\documentclass{aa}  
\usepackage[normalem]{ulem}
\usepackage{graphicx}
\usepackage{txfonts}
\usepackage{lipsum}
\usepackage{subcaption}         % necessary for continued figures, example in section 3
\usepackage{lscape}             % to rotate a single page table, example in appendix.
\usepackage{placeins}           % useful with \FloatBarrier, to keep 
\usepackage{amsmath}
\usepackage{tikz}
\usepackage{soul}
\usetikzlibrary{positioning, arrows.meta, fit, backgrounds}                                
\usepackage{hyperref}
\usepackage{color}
\usepackage{gensymb}

\newcommand\redsout{\bgroup\markoverwith{\textcolor{red}{\rule[0.5ex]{2pt}{0.4pt}}}\ULon}

\begin{document}

   \title{Geodesic-informed neural networks for testing spacetimes with S2}

   \author{Prajwal Hassan Puttasiddappa\inst{1, 2}\thanks{prajwal.puttasiddappa@edu.ufes.br}
          \and
          Dennis Fremstad\thanks{dennisfr@uio.no}\inst{1}
          \and 
          Abdolali Banihashemi\inst{1}
          \and
          David F. Mota\inst{1}
          }

   \institute{Institute of Theoretical Astrophysics, University of Oslo, Sem Sælands vei 13, 0371 Oslo, Norway\\
         \and
             PPGCosmo, Universidade Federal do Espírito Santo, 29075-910, Vitória, ES, Brazil
             }

   \date{\today}

% \abstract{}{}{}{}{} 
% 5 {} token are mandatory
 
  \abstract
  % context
  {Stellar orbits around Sagittarius~A$^*$, in particular that of S2, test the spacetime of the Galactic Center. Such analyses usually integrate the equations of motion for every likelihood evaluation and sample the posterior.}
  % aims
  {We develop a differentiable orbit model for static, spherically symmetric spacetimes that obtains the parameters, their uncertainties and the Bayesian evidence by optimization rather than sampling.}
  % methods
  {In the Geodesic-Informed Neural Network (GINN), the conserved quantities, the shape of the orbit and the redshift are computed exactly from the metric, and a neural network constrained by the geodesic equation represents the time evolution of the radial phase. The physical parameters and the network weights minimize a single loss with Adam and Gauss-Newton steps, and the curvature at the minimum gives the uncertainties and the evidence. We test this Gaussian approximation against a second-order expansion and an exact geodesic solver, and apply GINN to the S2 data in the Schwarzschild, Reissner-Nordstr\"om, hairy Horndeski, Bardeen, Janis-Newman-Winicour and Yukawa spacetimes.}
  % results
  {GINN reproduces the estimates and uncertainties of the exact solver. For Schwarzschild we find $M=(4.32\pm0.22)\times10^6\,M_\odot$ and $D=8.33\pm0.21$\,kpc. No spacetime deviates significantly from Schwarzschild; a Yukawa correction, for example, is limited to $|\delta|<0.011$ (95\%) at a range of about 180\,AU. No additional parameter improves the fit enough to pay its Occam penalty, so the Bayes factors range from inconclusive to moderate in favor of Schwarzschild. The Gaussian approximation is accurate when the hair enters the metric linearly and conservative when it enters only through its square.}
  % conclusions
  {When the posterior is close to Gaussian, GINN gives the estimates, their uncertainties and the evidence from a single optimization, and the same derivatives quantify departures from it.}
  
  % context heading (optional)
  % {} leave it empty if necessary  
   %{To investigate the physical nature of the `nuc\-leated instability' of
   %proto giant planets, the stability of layers
   %in static, radiative gas spheres is analysed on the basis of Baker's
   %standard one-zone model.}
  % aims heading (mandatory)
   %{It is shown that stability
   %depends only upon the equations of state, the opacities and the local
   %thermodynamic state in the layer. Stability and instability can
   %therefore be expressed in the form of stability equations of state
   %which are universal for a given composition.}
  % methods heading (mandatory)
   %{The stability equations of state are
   %calculated for solar composition and are displayed in the domain
   %$-14 \leq \lg \rho / \mathrm{[g\, cm^{-3}]} \leq 0 $,
   %$ 8.8 \leq \lg e / \mathrm{[erg\, g^{-1}]} \leq 17.7$. These displays
   %may be
   %used to determine the one-zone stability of layers in stellar
   %or planetary structure models by directly reading off the value of
   %the stability equations for the thermodynamic state of these layers,
   %specified by state quantities as density $\rho$, temperature $T$ or
   %specific internal energy $e$.
   %Regions of instability in the $(\rho,e)$-plane are described
   %and related to the underlying microphysical processes.}
  % results heading (mandatory)
   %{Vibrational instability is found to be a common phenomenon
   %at temperatures lower than the second He ionisation
   %zone. The $\kappa$-mechanism is widespread under `cool'
   %conditions.}
  % conclusions heading (optional), leave it empty if necessary 
   %{(tentative title)}

   \keywords{modified gravity--
                black hole --
                stellar orbits --
                physics-informed neural networks}

   \maketitle
%
%-------------------------------------------------------------------

\section{Introduction}

Scientific progress depends on the ability to extract physical information from observations. Ideally, one would infer physical laws directly from the data, with few or no assumptions about the underlying model. In practice, statistical and systematic uncertainties, instrumental limitations and sparse sampling make such a direct reconstruction difficult, and the usual task is more modest. One adopts a physical model and infers the parameter values that best describe the observations, along with their uncertainties. When the observables depend non-analytically on the parameters, the posterior is commonly explored by sampling with Markov chain Monte Carlo (MCMC) or nested sampling. Sampling is robust, but it requires a large number of evaluations of the forward model and does not use its derivatives. When the forward model is differentiable, gradient-based optimizers and samplers, and uncertainty estimates from the local curvature of the posterior, become available. In exoplanet studies, for example, differentiable orbit models have made gradient-based inference routine \citep{ForemanMackey2021, Thompson2023}.

Machine learning offers flexible, differentiable models of this kind for modelling. In particular, artificial neural networks (ANNs) can approximate complex relations between physical quantities and observables, and automatic differentiation \citep{Baydin2018} provides the derivatives of the network outputs with respect to all inputs and weights. In astrophysics and cosmology, however, data sets are often small, while ANNs typically need large training sets to converge. More importantly, an unconstrained ANN has no knowledge of the physical laws governing the system and can return unphysical predictions. Physics-informed neural networks (PINNs) address both problems by including the governing differential equations in the training objective, so that the network is constrained to physically allowed solutions \citep{PINN1, PINN2, Raissi2019}. In cosmology, PINNs have been trained to solve, for example, the background (Friedmann) equations over a range of parameters, so that they replace the numerical solver in fits to supernova, cosmic-chronometer and BAO data \citep{cosmoPINN}, to infer the dark-energy equation of state \citep{eosPINN}, and to reconstruct the Hubble function from supernova data \citep{HubblePINN}. In astrophysics, they have been used to solve the pulsar equation for axisymmetric, force-free pulsar magnetospheres \citep{astroPINN}.

Another application of PINNs, which is the focus of this work, is to infer the spacetime around the supermassive black hole at the center of our Galaxy, which can be directly probed by the orbits of the S-stars. Three decades of high-precision observations of the S-stars near Sagittarius~A$^*$ (Sgr~A$^*$) with the ESO New Technology Telescope, the Very Large Telescope (VLT) and the Keck Observatory have provided robust evidence for a compact object of about $4\times10^6\,M_\odot$ \citep{EckartGenzel1997, Ghez1998, Schodel2002, Ghez2008, Gillessen2009, Genzel2010SMBH, Gillessen2016data}, consistent with the horizon-scale images of the Event Horizon Telescope \citep{EHT2022SgrA}. Among the S-stars, S2 plays a central role, owing to its short orbital period ($\sim16$ yr) and high eccentricity ($e\sim0.88$). Around its 2018 pericenter passage, observations with the interferometer GRAVITY \citep{GRAVITY2017firstlight} and with Keck detected the combined gravitational redshift and transverse Doppler shift \citep{GRAVITY2018redshift, Do2019} and measured the distance to the Galactic Center to 0.3\% \citep{GRAVITY2019distance}. In 2020 the GRAVITY collaboration detected the Schwarzschild precession of about $12'$ per orbit, with $f_{\rm SP}=1.10\pm0.19$ ($f_{\rm SP}=1$ in GR), consistent with General Relativity (GR) \citep{GRAVITYSchprecess}, and later multi-star fits limited any extended mass inside the orbit of S2 \citep{GRAVITY2021DMdistribution1, GRAVITY2024extended}.

The astrometric and radial-velocity data of S2 have been used to test a wide range of modified-gravity theories \citep{scalartensorvector, scalartensorvectorMOG, BenistyDavis, fRS2, HorndeskiS2, beyondHorndeski, GRAVITYfifth, metricaffine2, R2gravity3}, alternative compact objects \citep{unveilingwormhole, nakedsingularity, RARold, RARrecent, ZakharovRAR}, and dark-matter distributions \citep{narrowbosonDM, ultralightbosnicDM, DMspike, GRAVITY2021DMdistribution1, DMsignature2, RARDM3} (see \citealt{deLaurentisreview2023} for a review). Most recently, several static black hole and horizonless spacetimes were examined in a Bayesian framework, with no significant preference for any of them over Schwarzschild \citep{testingbhspacetimebayesian, Bambhaniya:2026slo}. In these analyses the orbit is obtained by integrating the equations of motion numerically and the posterior is explored by sampling, as in GC-OrbitFit \citep{GCOrbitFit}, PyGRO \citep{pygro} and SNOPy \citep{prajLambda}. Each likelihood evaluation then requires a new numerical integration of the orbit, and the derivatives of the observables with respect to the parameters are not available.

In this work, we develop a differentiable forward model for the same problem, the Geodesic-Informed Neural Network (GINN). The conserved quantities, the shape of the orbit along with its precession, and the redshift are derived exactly from the metric at each point of the orbit, and the R{\o}mer delay is included in the projection onto the sky. What remains is when the star reaches each point, that is, its radial motion in time, $r(t)$. This is the relativistic generalization of Kepler's problem and has no closed-form solution. GINN represents it by a neural network built on the Keplerian solution and trained jointly on the data and on the geodesic equation of the metric under study. Every predicted position and velocity therefore depends on the network output, and the whole model is differentiable with respect to all orbital, reference-frame and metric parameters. The posterior can then be located by gradient-based minimization of a single loss and approximated by a Gaussian (Laplace approximation) with the network weights eliminated through their response to the parameters, instead of being sampled. GINN can be easily applied to any static, spherically symmetric metric, and here we apply it to six spacetimes. 

The paper is organized as follows. Section~\ref{sec:spacetimes} lists the spacetimes and their parameters. The data are described in Sec.~\ref{sec:data}. Section~\ref{sec:method} describes the orbit model, the network, the optimization and the uncertainty estimate. Section~\ref{sec:results} presents the results, and Sec.~\ref{sec:discussion} discusses their validation, and Sec.~\ref{sec:conclusion} concludes. The appendices give details of the optimization, the evidence, the higher-order checks and the validation tests.

\section{Spacetimes}
\label{sec:spacetimes}

We consider static, spherically symmetric spacetimes,
\begin{equation}\label{eq:line_element}
    ds^2 = -f(r)\,dt^2 + g(r)\,dr^2 + h(r)\,\big(d\vartheta^2 + \sin^2\vartheta\,d\varphi^2\big),
\end{equation}
in geometric units ($G=c=1$). The spin of Sgr~A$^*$ affects the orbit of S2 at higher order than the Schwarzschild precession and is below the sensitivity of the available data \citep{SgrAspin2025}. Each spacetime is characterized by the mass $M$ and, except for Schwarzschild, one additional parameter $\xi$, which we call the hair parameter. All of them reduce to Schwarzschild for vanishing hair (the Yukawa metric to its weak-field form). We follow a selection of recent MCMC studies \citep{testingbhspacetimebayesian, Bambhaniya:2026slo} so that the results can be compared directly. A spacetime enters the analysis only through the functions $f$, $g$ and $h$ in the forward model of Sec.~\ref{sec:method}. 

Since S2 never comes closer than its pericenter $r_p\simeq2800\,M$, the observations probe only the exterior spacetime. Whether a horizon, a regular core or a singularity lies inside is not tested. We also treat the region outside the central object as empty, apart from the fields of the theory itself. This is supported by the upper limit of $1200\,M_\odot \approx 3 \times 10^{-4}M$ (at $1\sigma$ confidence) on any extended mass, such as a stellar cusp or a dark-matter spike, inside the orbit of S2 \citep{GRAVITY2024extended}. In GR, Birkhoff's theorem and its electrovacuum extension make the Schwarzschild and RN metrics the exterior of any spherical distribution of mass or charge. For the other theories, no such uniqueness result is available, so our analysis tests the specific metrics rather than every object these theories allow.

\subsection{Schwarzschild spacetime}
The unique spherically symmetric vacuum solution of GR \citep{Schwarzschild1916},
\begin{equation}
    f(r) = 1 - \frac{2M}{r}, \qquad g(r) = f(r)^{-1}, \qquad h(r) = r^2 ,
\end{equation}
is the reference model. It has a horizon at $r=2M$, an unstable circular photon orbit at $r=3M$ and an innermost stable circular orbit at $r=6M$. The pericenter of S2, $r_p\simeq2800\,M$, lies far outside this strong-field region, so S2 probes the weak field, where $M/r_p \simeq 4 \times 10^{-4}$ sets the size of the relativistic effects in the orbit.

\subsection{Reissner-Nordstr\"om spacetime}
The exterior of a spherical mass with electric charge $Q$ \citep{Reissner1916, Nordstrom1918} is
\begin{equation}
    f(r) = 1 - \frac{2M}{r} + \frac{Q^2}{r^2}, \qquad g(r) = f(r)^{-1}, \qquad h(r) = r^2 ,
\end{equation}
with hair parameter $q_{\rm RN}=Q/M$. It contains a black hole for $|q_{\rm RN}|\le1$ and a naked singularity for $|q_{\rm RN}|>1$. The charge enters only through $q_{\rm RN}^2$, so its sign cannot be measured. In the radial equation (see~\eqref{eq:radial}), the $Q^2/r^2$ term in $f$ adds to the centrifugal term $L^2/r^2$, so for given turning points its leading effect on the radial motion is absorbed by $L$ (Fig.~\ref{fig:V}). The electric charge of Sgr~A$^*$ is expected to be negligible \citep{Zajacek2018}, so a bound on $q_{\rm RN}$ is best read as a bound on any positive $1/r^2$ term in $f$. The same metric with a coefficient $\beta$ of either sign in place of $Q^2$ describes braneworld black holes with tidal charge \citep{Dadhich2000}, whereas our parametrization covers only $\beta=Q^2\ge0$.

\subsection{Hairy Horndeski spacetime}
Horndeski gravity is the most general scalar-tensor gravity with second-order field equations \citep{Horndeski1974}. Its quartic sector admits the exact hairy solution \citep{Bergliaffa2021}
\begin{equation}
    f(r) = 1 - \frac{2M}{r} + \frac{Q_H}{r}\ln\!\left(\frac{r}{2M}\right), \quad g(r) = f(r)^{-1}, \quad h(r) = r^2 ,
\end{equation}
with hair parameter $q_H=Q_H/M$. For $q_H>-2$, which includes all values considered here, the horizon stays at $r=2M$. The hair modifies the Newtonian potential itself, with $m(r)=M-(Q_H/2)\ln(r/2M)$. Since $m(r)$ has no finite limit as $r\to\infty$, $M$ is the parameter that fixes the horizon rather than an asymptotic mass. Over the orbit of S2 the logarithm grows from about 7 at pericenter to about 10 at apocenter, so the star feels a different mass at the two ends of its orbit. The constant part of this shift is indistinguishable from a change of $M$, which makes $q_H$ partly degenerate with the mass (and sign of $q_H$ matters).

\subsection{Bardeen spacetime}
The Bardeen spacetime \citep{Bardeen1968} replaces the central singularity with a regular core and can be interpreted as a magnetic monopole of charge $Q_B$ in nonlinear electrodynamics \citep{AyonBeato2000}. In that case, the metric coefficients read as
\begin{equation}
    f(r) = 1 - \frac{2Mr^2}{(r^2 + Q_B^2)^{3/2}}, \qquad g(r) = f(r)^{-1}, \qquad h(r) = r^2 ,
\end{equation}
with hair parameter $q_B=Q_B/M$. Its mass function $m(r)=Mr^3/(r^2+Q_B^2)^{3/2}$ vanishes at the center, which is a regular de Sitter core, and approaches $M$ for $r\gg Q_B$. It has horizons for $|q_B|\le4/(3\sqrt3)\simeq0.77$ and is a horizonless regular object otherwise. At large radii $m(r)\simeq M[1-3Q_B^2/(2r^2)]$, so the charge enters only through $q_B^2$, and at the pericenter of S2 the relative change of the mass is about $2\times10^{-7}q_B^2$. Only $|q_B|\gg1$ can therefore affect the orbit, which lies deep in the horizonless regime.

\subsection{Janis-Newman-Winicour (JNW) spacetime}
The JNW solution describes a mass with a minimally coupled\footnote{Static, spherically symmetric, vacuum solutions of Brans-Dicke, like \citep{Brans:1962zz, HassanPuttasiddappaBD} become JNW in the Einstein frame \citep{Bhadra:2006rn}.}, massless scalar field of charge $Q_{\rm JNW}$ \citep{JNW1968}. In the notation of \citet{Virbhadra1997},
\begin{equation}
    f(r) = g(r)^{-1} = \left(1 - \frac{b}{r}\right)^{\gamma}, \qquad
    h(r) = r^2\left(1 - \frac{b}{r}\right)^{1-\gamma},
\end{equation}
with $b=2\sqrt{M^2+Q_{\rm JNW}^2}$, $\gamma=2M/b$ and hair parameter $q_{\rm JNW}=Q_{\rm JNW}/M$. For $q_{\rm JNW}\neq0$, the surface $r=b$ is a naked curvature singularity, and a photon sphere exists only for $|q_{\rm JNW}|<\sqrt3$ \citep{VirbhadraEllis2002}. Since $r$ is not the areal radius $R=\sqrt h$, we expand in $R$: $f=1-2M/R+MQ_{\rm JNW}^2/(3R^3)+\dots$ and $g_{RR}\equiv g(r)\,({\rm d}r/{\rm d}R)^2=1+2M/R+(4M^2-Q_{\rm JNW}^2)/R^2+\dots$. The charge enters only through $q_{\rm JNW}^2$, with corrections to Schwarzschild suppressed by $Q_{\rm JNW}^2/R^2\sim10^{-7}q_{\rm JNW}^2$ at the radii of S2. Only $|q_{\rm JNW}|\gg1$ can therefore affect the orbit, lying far outside the photon sphere regime. Thus, S2 and horizon-scale imaging probe different regions of the JNW parameter space. 

\subsection{Yukawa-type correction}
In $f(R)$ gravity and other theories with a massive scalar degree of freedom, the weak-field potential acquires a Yukawa term \citep{Yukawa1, Yukawa2, fRS2},
\begin{equation}
\begin{split}
    f(r) = 1 + \Phi(r), &\qquad g(r) = 1 - \Phi(r), \qquad h(r) = r^2,\\
    \Phi(r) &= -\frac{2M}{(1+\delta)\,r}\left(1 + \delta\, e^{-r/\lambda}\right),
\end{split}
\end{equation}
with strength $\delta$ and range $\lambda$, the Compton wavelength of the extra degree of freedom. The mass function $m(r)=M(1+\delta e^{-r/\lambda})/(1+\delta)$ equals $M$ for $r\ll\lambda$ and $M/(1+\delta)$ for $r\gg\lambda$. For $\lambda$ much larger than the orbit,
\begin{equation}
    \Phi(r)\simeq-\frac{2M}{r}+\frac{2\delta M}{(1+\delta)\lambda}-\frac{\delta M\,r}{(1+\delta)\lambda^2}.
\end{equation}
The constant term does not affect the Newtonian motion and only shifts the redshift by a constant, which is absorbed by the radial-velocity offset (Sec.~\ref{sec:method}). The last term is a uniform radial force whose strength depends only on $\delta/[(1+\delta)\lambda^2]$. Since $\delta$ and $\lambda$ cannot be estimated separately in this regime, we fix $\lambda$ and take $\xi\equiv\delta$ (Sec.~\ref{sec:results}). Yukawa-type corrections have previously been constrained with S-stars by \citet{Borka2013, Hees2017GRtests, fRS2, GRAVITYfifth, Jovanovic2026fifth}.

\section{Data}\label{sec:data}

We use the public astrometric and spectroscopic data of S2 from \citet{Gillessen2016data}, obtained from the catalog of \citet{catalogue}. The dataset consists of 145 astrometric positions and 44 line-of-sight radial velocity measurements between 1992 and 2016, that is, $N = 2\times145+44=334$ data points. The astrometry combines speckle imaging with the SHARP camera at the ESO New Technology Telescope (NTT) before 2002 with adaptive-optics imaging with NACO at the Very Large Telescope (VLT) from 2002 onward. The radial velocities come mostly from the integral-field spectrograph SINFONI at the VLT, complemented by earlier measurements (before 2003) with NIRC2 at the Keck Observatory in slit-spectroscopy mode. We take all uncertainties as published and treat them as independent and Gaussian. %Unrecognized confusion with faint stars can add time-correlated astrometric errors at the sub-milliarcsecond level, in particular near pericenter \citep{PlewaSari2018}, and we do not model this.

The positions of S2 are measured in near-infrared images relative to an astrometric reference frame tied to the radio frame through SiO maser stars (evolved stars detected at both infrared and radio wavelengths) \citep{Plewaoffset}. The orbit of S2 is centered on the dynamical center, whose position and motion relative to the frame origin are known only to the accuracy of this calibration. We therefore fit five offset parameters together with the orbit: the position $(x_0,y_0)$ and proper motion $(v_{x,0},v_{y,0})$ of the dynamical center, and a constant offset $v_{z,0}$ of the radial velocities. The first four have Gaussian priors of zero mean and widths $0.2\,\mathrm{mas}$ (at epoch 2009.0) and $0.1\,\mathrm{mas\,yr^{-1}}$ from this calibration, as in \citet{Gillessen2016data}; $v_{z,0}$ has no prior.

\section{Method}\label{sec:method}

\begin{figure}
\centering
\begin{tikzpicture}[
  font=\footnotesize, node distance=4mm and 3mm, >={Stealth[length=1.6mm]},
  box/.style={draw, rounded corners=2pt, align=center, inner sep=3pt, minimum width=#1, minimum height=7mm},
  box/.default=4.2cm,
  exact/.style={box=#1, fill=blue!6},     exact/.default=4.2cm,
  nn/.style={box=#1, fill=orange!12},     nn/.default=4.2cm,
  opt/.style={box=#1, fill=green!8},      opt/.default=2.1cm,
  out/.style={box=#1, fill=gray!10},      out/.default=3.3cm,
  arr/.style={->, thick},
  lab/.style={font=\scriptsize\itshape, fill=white, inner sep=1pt}]

% ---- left column: model chain; right column: data and physics residual
\node[box] (metric) {\textbf{Metric} $f,\,g,\,h$, parameters\\ $\Theta=\{M,D,a,e,i,\Omega,\omega,t_p$, offsets, $\xi\}$};

\node[box=2.9cm, right=of metric] (data) {\textbf{Data} (S2, 1992-2016)\\ 145 positions, 44 RVs};

\node[exact, below=of metric] (geo) {\textbf{Exact geodesic quantities}\\ $E,L$ from $r_p,r_a$\\ $d\phi/dt=F(\phi;\Theta)$\\ $n_r=2\pi/P_r$,\ $\varphi(\phi)$};

\node[nn, below=6mm of geo] (ginn) {\textbf{GINN}\\ $\phi(t)=\phi_K\big(n_r(t-t_p)\big) + \phi_{\rm NN}\ ,$ \\$\phi_{\rm NN} \equiv \mathcal N_\theta(\phi_K)-\mathcal N_\theta(0)$};

\node[nn=2.9cm, right=5mm of ginn] (res) {\textbf{Physics residual}\\ $R_c=(\dot\phi_{\rm NN}-F)/F$};

\node[exact, below=6mm of ginn] (obs) {\textbf{Observables}\\ $r(\phi),\varphi(\phi)\to$ sky projection,\\ R{\o}mer delay, redshift, offsets\\ $\to$ RA, Dec, $v_{\rm LOS}$};

% ---- loss (spans both columns)
\node[box=7.2cm, fill=red!6, below=5mm of obs.south west, anchor=north west] (loss)
  {\textbf{Joint loss}\\ $\mathcal L=\chi^2_{\rm data}+\chi^2_{\rm frame}+\sum_c (R_c/\epsilon)^2+\lVert\theta\rVert^2/\sigma_w^2\;$};

% ---- optimisation, three stages in a row
\node[opt=1.9cm, below=6mm of loss.south west, anchor=north west] (s1) {\textbf{Pre-fit}\\ Kepler orbit};
\node[opt=1.6cm, right=of s1] (s2) {\textbf{Adam}\\ $(\mathbf u,\theta)$};
\node[opt=2.9cm, right=of s2] (s3) {\textbf{Gauss-Newton}\\ $\Delta\mathbf u=-H_m^{-1}\mathbf g_m$};

% ---- outputs
\node[out, below=10mm of s3.south east, anchor=north east] (lap) {\textbf{Laplace errors}\\ $\Sigma_u=(\hat J^\top\hat J)^{-1}$\\ {\scriptsize (network follows $\mathbf{u}$)}};
\node[out=3.0cm, left=of lap] (est) {\textbf{Estimates} $\Theta^\ast$\\ samples, corner plots};

% ---- arrows
\draw[arr] (metric) -- (geo);
\draw[arr] (geo) -- node[lab, right] {$n_r$, $e$} (ginn);
\draw[arr] (geo.east) -| node[lab, pos=0.75, right] {$F(\phi)$} (res.north);
\draw[arr] (ginn) -- node[lab, above] {$\dot\phi_{\rm NN}$} (res);
\draw[arr] (ginn) -- node[lab, right] {$\phi(t)$} (obs);
\draw[arr] (obs.south) -- (obs.south |- loss.north);
\draw[arr] (res.south) -- (res.south |- loss.north);
\draw[arr] (data.east) -- ++(2.5mm,0) |- ([xshift=-2mm, yshift=2.5mm]loss.north east) -- ([xshift=-2mm]loss.north east);
\draw[arr] (loss.south -| s2.north) -- (s2.north);
\draw[arr] (s1) -- (s2);
\draw[arr] (s2) -- (s3);
\draw[arr, dashed, gray] ([xshift=4mm]s3.south west) -- ++(0,-4mm) -|
      node[lab, pos=0.25, below, text=gray!70!black] {re-train} (s2.south);
\draw[arr] (s3.south -| lap.north) -- node[lab, right] {converged} (lap.north);
\draw[arr] (lap) -- (est);

% ---- legend
\node[below=3mm of est.south west, anchor=north west, font=\scriptsize, align=left]
  {\tikz\node[fill=blue!6,draw,minimum size=2.2mm]{}; exact (metric)\quad
   \tikz\node[fill=orange!12,draw,minimum size=2.2mm]{}; network\quad
   \tikz\node[fill=green!8,draw,minimum size=2.2mm]{}; optimisation\quad
   \tikz\node[fill=gray!10,draw,minimum size=2.2mm]{}; output};
\end{tikzpicture}
\caption{Workflow of GINN. Everything that can be computed exactly from the metric (blue) is computed exactly; the network (orange) represents only the radial phase $\phi(t)$ and is constrained by the geodesic equation through the physics residual. All parameters minimize one joint loss in three stages (green: Keplerian pre-fit, Adam, Gauss-Newton); the curvature of the data and frame-prior residuals at the minimum, with the network following the parameters, gives the Laplace uncertainties.}
\label{fig:workflow}
\end{figure}
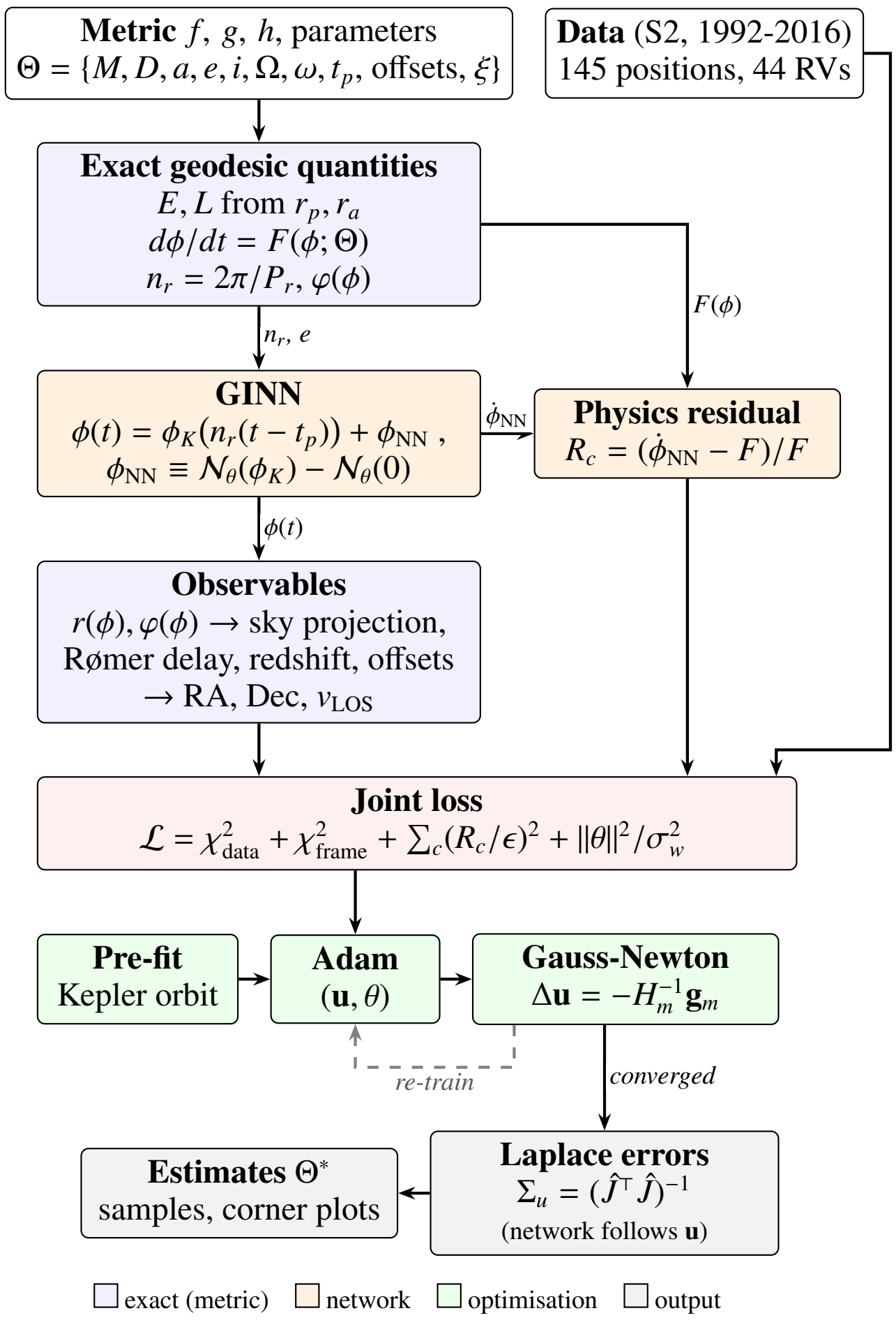

Figure~\ref{fig:workflow} summarizes the method, whose steps are described in the following subsections.

%%%%%%%%%%%%%%%%%%%%%

\subsection{Orbit modeling}

The orbit of S2 is described by the six orbital elements: semi-major axis $a$, eccentricity $e$, inclination $i$, longitude of the ascending node $\Omega$, argument of pericenter $\omega$ and time of pericenter passage $t_p$. The spacetime is fixed by the mass $M$ and the hair parameter of Sec.~\ref{sec:spacetimes},
\begin{equation}
    \xi \in \{\,q_{\rm RN},\ q_H,\ q_B,\ q_{\rm JNW},\ \delta\,\} .
\end{equation}
Distances are converted to angles with the distance $D$ to the Galactic Center. The parameters $M$, $a$ and $D$ must be positive, and a bound orbit requires $0<e<1$. We therefore work with the transformed variables $\ln (M/M_\odot)$, $\ln (a/{\rm AU})$, $\ln(D/{\rm kpc})$ and $\mathrm{logit}\,e=\ln[e/(1-e)]$. The constraints are then satisfied automatically, without bounds in the optimization, and all parameters have a similar numerical range, which improves the conditioning of the optimization. Together with the offset parameters (Sec.~\ref{sec:data}), with $(x_0,y_0)$ given at the reference epoch $t_{\rm ref}$, the complete parameter vector is
\begin{equation}
    \Theta = \{\ln M, \ln a, \mathrm{logit}\, e, i,\Omega, \omega, t_p, \ln D, x_0, y_0, v_{x,0}, v_{y,0}, v_{z,0}, \xi \}\ .
\end{equation}
Spherical symmetry confines a geodesic to a plane, which we take as $\vartheta=\pi/2$. Time-translation and rotational invariance of Eq.~\eqref{eq:line_element} give two conserved quantities per unit rest mass, the specific energy $E = f(r)\,\dot{t}$ and angular momentum $L = h(r)\,\dot{\varphi}$, where dots denote derivatives with respect to proper time $\tau$. The normalization $g_{\mu\nu}\dot{x}^\mu \dot{x}^\nu = -1$ leads to the radial equation of motion,
\begin{equation}
  g(r)\left(\frac{dr}{d\tau}\right)^2 = \frac{E^2}{f(r)} - 1 - \frac{L^2}{h(r)} \equiv V(r).
  \label{eq:radial}
\end{equation}
A bound orbit oscillates between the pericenter $r_p=a(1-e)$ and the apocenter\footnote{For JNW, $r$ is not the areal radius. We therefore write the metric in the areal radius $R=\sqrt h$, so that all quantities of this section, including $a$, $e$, the turning points and the anomaly $\phi$, are defined in $R$. For the other spacetimes $h=r^2$ and nothing changes.} $r_a=a(1+e)$, where $V(r_p)=V(r_a)=0$. These two conditions are linear in $E^2$ and $L^2$, 
\begin{equation}
  \frac{E^2}{f(r_a)} - \frac{L^2}{h(r_a)} = 1, \qquad
  \frac{E^2}{f(r_p)} - \frac{L^2}{h(r_p)} = 1\ ,
  \label{eq:EL}
\end{equation}
and determine $E$ and $L$ uniquely. The elements $a$ and $e$ are thus defined through the turning points of the areal radius, which coincides with $r$ for $h = r^2$. 

\begin{figure}
  \centering
  \includegraphics[width=\columnwidth]{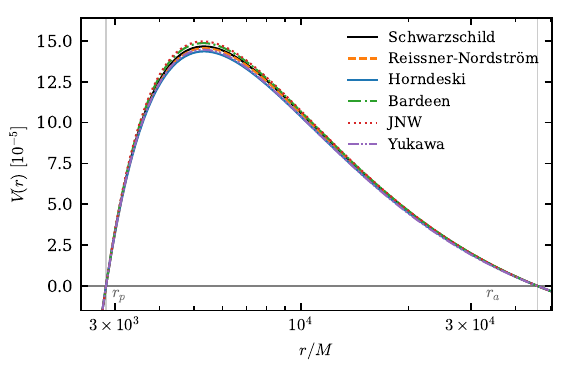}
  \caption{Radial function $V(r)$ of Eq.~\eqref{eq:radial} for the S2 orbit in each spacetime, with the turning points $r_p$ and $r_a$ marked by the vertical lines. The hair parameters are chosen for illustrative purposes to distinguish the curves.}
  \label{fig:V}
\end{figure}

Figure~\ref{fig:V} compares the radial potential $V(r)$ across the spacetimes of Sec.~\ref{sec:spacetimes}, with $E$ and $L$ fixed by the common turning points. The shape of $V$ between the turning points sets the radial speed, and $g$ rescales it in Eq.~\eqref{eq:radial} without explicitly entering $V(r)$.

Instead of solving for $r(\tau)$ directly, we introduce the relativistic anomaly $\phi$ \citep{Darwin1959, Cutler1994},
\begin{equation}
  r(\phi) \equiv \frac{p}{1+e\cos\phi}, \qquad p \equiv a(1-e^2)\ .
  \label{eq:rchi}
\end{equation}
This is an exact relabeling of the radial coordinate between the two turning points, valid for any metric, and in Newtonian gravity $\phi$ is the true anomaly. With $dt/d\tau=E/f$, Eq.~\eqref{eq:radial} gives the rate of $\phi$ in coordinate time,
\begin{equation}\label{eq:dchi_dt}
    \frac{d\phi}{dt} = \frac{f(r)}{E}\,\sqrt{\frac{V(r)}{g(r)}}\,\left|\frac{dr(\phi)}{d\phi}\right|^{-1} \equiv F(\phi;\Theta)\ ,
\end{equation}
which, for given parameters, is a closed-form function of $\phi$. Both $V$ and $dr/d\phi$ vanish at the turning points while their ratio remains finite, so we evaluate $V$ relative to the nearest turning point where the leading constant terms cancel exactly. The radial period and the radial frequency follow from the rate,
\begin{equation}
    P_r =\int_0^{2\pi} \frac{d\phi}{F(\phi; \Theta)}\ ,\qquad  n_r=\frac{2\pi}{P_r}.
\label{eq:nr}
\end{equation}
Similarly, conservation of specific angular momentum $\mathrm{d}\varphi/\mathrm{d}\tau = L/h(r)$, gives the shape of the orbit,
\begin{equation}\label{eq:dphi_dchi}
  \varphi(\phi) = \int_0^{\phi}\frac{L/h\big(r(\phi')\big)}{d\phi'/d\tau}\,d\phi'\ .
\end{equation}
The pericenter advance per radial period is then $\Delta\varphi = \varphi(2\pi) - 2\pi$. In Newtonian gravity (where $\varphi=\phi$), $\Delta\varphi=0$, and in the Schwarzschild case $\Delta\varphi \simeq 6\pi M/p$ to leading order. We evaluate both integrals \eqref{eq:nr} and \eqref{eq:dphi_dchi} with Gauss-Legendre quadrature.

The shape of the orbit, including its precession, is therefore known exactly from the metric. What remains is its time dependence $\phi(t)$, i.e. the inverse of $t(\phi)=\int F^{-1}d\phi$. Even in Newtonian gravity, where this inversion is Kepler's equation, $\mathcal E-e\sin\mathcal E=n(t-t_p)$ for the eccentric anomaly $\mathcal E$, has no closed-form solution and is strongly non-linear at the eccentricity of S2.  This is the part of the problem that we represent with a neural network (Sec.~\ref{sec:ginn}).

\subsection{From $\phi(t)$ to the observables}\label{sec:observables}

Once $\phi(t)$ is known, the observables follow from the following steps.

\begin{enumerate}
    \item At each epoch we compute the areal radius $R=\sqrt{h(r)}$ and the azimuth $\varphi(\phi)$, and hence the position coordinates $X=R\cos\varphi$, $Y=R\sin\varphi$ in the orbital plane, with $X$ pointing to pericenter. The velocities follow from the conserved quantities,
    \begin{equation}
        v_R=\pm\frac{dR}{dr}\,\frac{\sqrt{V(r)/g(r)}}{dt/d\tau},\qquad
        v_\varphi=R\,\frac{L/h(r)}{dt/d\tau},
    \end{equation}
    with $dR/dr=h'/(2\sqrt h)$, which equals one for $h=r^2$, and $v_R>0$ while the star moves outward ($\sin\phi>0$). In Cartesian coordinates, $v_X=v_R\cos\varphi-v_\varphi\sin\varphi$ and $v_Y= v_R\sin\varphi+v_\varphi\cos\varphi$.

    \item The orbital plane is projected on the sky with the unit vectors of the $X$ and $Y$ axes,
    \begin{gather}
    \mathbf e_X=\begin{pmatrix}\cos\Omega\cos\omega-\sin\Omega\sin\omega\cos i\\ \sin\Omega\cos\omega+\cos\Omega\sin\omega\cos i\\ \sin\omega\sin i\end{pmatrix},\nonumber\\
    \mathbf e_Y=\begin{pmatrix}-\cos\Omega\sin\omega-\sin\Omega\cos\omega\cos i\\ -\sin\Omega\sin\omega+\cos\Omega\cos\omega\cos i\\ \cos\omega\sin i\end{pmatrix},
    \end{gather}
    whose components are the Thiele-Innes constants, so that
    \begin{equation}
    \begin{split}
        (x_{\rm Dec},\,x_{\rm RA},\,Z)^\top &= X\,\mathbf e_X+Y\,\mathbf e_Y,\\
        v_z &= v_X\,(\mathbf e_X)_3+v_Y\,(\mathbf e_Y)_3 ,
    \end{split}
    \end{equation}
    where $Z$ and $v_z$ are positive away from the observer.

    \item Light received at time $t_{\rm obs}$ was emitted at $t_{\rm em}=t_{\rm obs}-Z(t_{\rm em})/c$. We account for the R{\o}mer delay between emission and reception. Identifying $t$ with the distant-observer time, up to a constant absorbed into $t_p$, we solve this relation with one fixed-point iteration, $t_{\rm em}\simeq t_{\rm obs}-\frac{Z(t_{\rm obs})}{c}$ whose timing error is of order $|Z v_z|/c^2$, below one hour for S2. We evaluate the orbital quantities at $\phi(t_{\rm em})$.
    
    \item The positions are converted to angles with $D$, and the frame offsets are added, 
    \begin{equation}
    \begin{split}
        \mathrm{RA}^{\rm mod}(t_{\rm obs})&=\frac{x_{\rm RA}}{D}+x_0+v_{x,0}\,(t_{\rm obs}-t_{\rm ref}),\\
        \mathrm{Dec}^{\rm mod}(t_{\rm obs})&=\frac{x_{\rm Dec}}{D}+y_0+v_{y,0}\,(t_{\rm obs}-t_{\rm ref}),
    \end{split}
    \end{equation}
    where $t_{\rm ref}$ is the mean astrometric epoch. The line-of-sight velocity follows from the frequency shift seen by a static observer at infinity,
    \begin{equation}
        1+z=\frac{dt}{d\tau}\,(1+v_z)=\frac{E}{f(r)}\,(1+v_z),\qquad v^{\rm mod}_{\rm LOS} = c\,z+v_{z,0}\ .
    \end{equation} 
    The factor $E/f$ contains the gravitational redshift and the transverse Doppler shift, and $(1+v_z)$ the longitudinal Doppler shift. Since both are evaluated with the metric under study, the redshift is consistent with each spacetime. We neglect the Shapiro delay, which is below the precision of these data.
\end{enumerate}

\subsection{Geodesic-informed neural network (GINN)}\label{sec:ginn}

We write the relativistic anomaly as a Keplerian part plus a neural-network correction,
\begin{equation}
  \phi(t)=\phi_K\!\big(n_r(t-t_p);e\big)+\mathcal N_\theta(\phi_K)-\mathcal N_\theta(0) \ . 
  \label{eq:ginn}
\end{equation}
The Keplerian part $\phi_K$ is the true anomaly obtained from Kepler's equation with the mean anomaly $n_r(t-t_p)$, where $n_r$ is the exact radial frequency of the metric, Eq.~\eqref{eq:nr}, so that $\phi_K$ already has the correct period. The network $\mathcal N_\theta$, with weights $\theta$, is fully connected, takes $(\cos\phi_K,\sin\phi_K)$ as input, has two hidden layers of 32 $\tanh$ units and a linear output scaled by $10^{-3}$\,rad, about $1200$ weights in total. Because its input is periodic, the correction repeats every radial period and cannot produce a secular drift. Subtracting $\mathcal N_\theta(0)$ makes the correction vanish at pericenter, so that $t_p$ keeps its meaning and is not degenerate with a constant network output. Since the precession is already carried by $\varphi(\phi)$, the network represents the small, periodic, non-Keplerian part of the radial motion, $\phi(t)-\phi_K(t)$.

The network is tied to the equation of motion~\eqref{eq:dchi_dt} through the relative residual
\begin{equation}\label{eq:residual}
\begin{split}
    R_c&=\frac{\dot\phi_{\rm NN}-F(\phi;\Theta)}{F(\phi;\Theta)}\ ,\\
    {\rm with, }\quad  \dot\phi_{\rm NN}&=\big[1+\mathcal N_\theta'(\phi_K)\big]\,\frac{n_r(1+e\cos\phi_K)^2}{(1-e^2)^{3/2}}\ ,
\end{split}
\end{equation}
where $\phi$ is the anomaly of Eq.~\eqref{eq:ginn} and the last factor is the Keplerian rate $\dot\phi_K$. The derivative $\mathcal N_\theta'$ is computed exactly by automatic differentiation \citep{Baydin2018}. We evaluate $R_c$ at $N_c=256$ collocation phases, spaced uniformly in the Keplerian eccentric anomaly over one radial period and shifted together by a random amount at each training step. The random shift enforces the equation of motion at all phases rather than at a fixed set of points. This stochastic resampling is related to sampling strategies used in PINNs \citep{Wu2023PINNsampling}. Because $R_c$ is a relative error of the rate, it has the same weight near pericenter, where $\phi$ changes fast, and near apocenter, where it changes slowly.

%Because $R_c$ is a relative error of the rate, it has the same meaning at pericenter, where $\phi$ changes fast, and at apocenter, where it changes slowly.
%\den{Make sure that the text reflexts the model. For instance, I am not sure if we actually shuffle the data during training.}\praj{we are not shuffling the data. We are shuffling the collocation points - this is where we compute the physics residuals. We shuffle it because we do not want to overfit at particular 256 points - its kinda like regularization. By stochastically jittering the points we are making sure that physics is valid in any random point.}

Unlike a standard PINN, which represents the full solution, GINN computes exactly everything that follows from the metric ($E$, $L$, $n_r$, $\varphi(\phi)$ and the redshift) and leaves only $\phi(t)-\phi_K(t)$ to the network. For S2, this correction is about $10^{-4}$\,rad in Schwarzschild, or $\lesssim0.02$\,mas on the sky, below the present astrometric precision. The network therefore adds no information beyond an exact solution of the equation of motion. Its role is to make the forward model differentiable with respect to all parameters, so that they can be estimated jointly by gradient-based optimization. The same construction applies to metrics for which inverting $t(\phi)$ is inconvenient, and to more precise data, for which the correction becomes measurable.

\subsection{Loss and priors} \label{subsec:loss}

All parameters are obtained by minimizing a single loss, 
\begin{equation}
  \mathcal L(\Theta,\theta)=\chi^2_{\rm data} + \chi^2_{\rm frame}
  +\sum_{c=1}^{N_c}\left(\frac{R_c}{\epsilon}\right)^2
  +\frac{\lVert\theta\rVert^2}{\sigma_w^2}\ ,
  \label{eq:loss}
\end{equation}
with
\begin{multline}
  \chi^2_{\rm data}=\sum_{i}\left[\left(\frac{\mathrm{RA}^{\rm mod}_i-\mathrm{RA}_i}{\sigma_{{\rm RA},i}}\right)^2
  +\left(\frac{\mathrm{Dec}^{\rm mod}_i-\mathrm{Dec}_i}{\sigma_{{\rm Dec},i}}\right)^2\right]\\
  +\sum_j\left(\frac{v^{\rm mod}_{{\rm LOS},j}-v_{{\rm LOS},j}}{\sigma_{v,j}}\right)^2\ ,
\end{multline}
where $i$ runs over the astrometric epochs and $j$ over the radial-velocity epochs. The Gaussian frame priors of Sec.~\ref{sec:data} enter as $\chi^2_{\rm frame}$, the sum of four squared residuals: the position of the dynamical center at epoch 2009.0, $x_0+v_{x,0}(2009.0-t_{\rm ref})$ and likewise in $y$, divided by $0.2$\,mas; and $v_{x,0}$ and $v_{y,0}$ divided by $0.1\,{\rm mas\,yr^{-1}}$. They are treated as four additional data residuals throughout, in the optimization, the uncertainties and the evidence. The physics term has weight $1/\epsilon^2$. It enforces the geodesic equation up to a relative tolerance $\epsilon$, analogous to the tolerance set in an ODE solver. We set $\epsilon=10^{-4}$ by a convergence test: for $10^{-4} \lesssim \epsilon \lesssim10^{-3}$ all results agree with an exact geodesic solver and do not depend on $\epsilon$. A loose tolerance lets the network adapt to noise, and a tighter one makes the problem too stiff to optimize (Appendix~\ref{app:validation}). The last term is a weak Gaussian prior on the network weights ($\sigma_w=10$) \citep{MacKay1992}. It has no noticeable effect on the fit, but makes the curvature matrix invertible: the network has more weights than there are residuals ($334+4+256=594<1200$), so without it the curvature of the loss (the Gauss-Newton matrix) is singular in the weight directions, and the prior adds $\sigma_w^{-2}$ to its diagonal (see Sec.~\ref{sec:optimization}).

The hair parameters have the uniform priors of Table~\ref{tab:priors}. A uniform prior is constant inside its interval and adds nothing to the loss or its curvature. We enforce the bounds by projection; that is, an update that leaves the interval is set back to the boundary. The two kinds of prior therefore act differently. A uniform prior is a hard wall that is felt only when a step reaches it, whereas a Gaussian frame prior is soft: it adds to the loss and its curvature everywhere and pulls the offsets towards zero with a strength set by its width, so that values beyond the width are allowed but penalized. Apart from the frame priors, the mass, the distance, the orbital elements and $v_{z,0}$ have no prior term in the loss; that is, their priors are flat in the fitted variables of $\Theta$. Since these parameters are common to all spacetimes, their priors cancel in the Bayes factors.

\begin{table}
  \centering
  \caption{Hair parameters and their uniform priors. For $q_{\rm RN}$, $q_B$ and $q_{\rm JNW}$ the likelihood depends only on $\xi^2$, so $\xi$ and $-\xi$ describe the same spacetime. We nevertheless use a prior symmetric about zero rather than, e.g., $\mathcal U[0,50]$: both give the same posterior for $|\xi|$ and the same evidence, but the symmetric prior keeps the maximum at $\xi=0$ inside the prior rather than on its boundary, where the Laplace approximation would not apply. For $q_H$ and $\delta$ the sign is physical.}
  \label{tab:priors}
  \begin{tabular}{lll}
    \hline\hline
    Spacetime & Hair parameter $\xi$ & Prior \\
    \hline
    Reissner-Nordstr\"om & $q_{\rm RN}=Q/M$ & $\mathcal U[-50,50]$ \\
    Hairy Horndeski & $q_H=Q_H/M$ & $\mathcal U[-0.05,0.05]$ \\
    Bardeen & $q_B=Q_B/M$ & $\mathcal U[-200,200]$ \\
    JNW & $q_{\rm JNW}=Q_{\rm JNW}/M$ & $\mathcal U[-600,600]$ \\
    Yukawa & $\delta$ (at fixed $\lambda$) & $\mathcal U[-0.5,0.5]$ \\
    \hline
  \end{tabular}
\end{table}

\subsection{Optimization}\label{sec:optimization}

The optimization has to deliver two things: the minimum of the loss, which is the maximum of the posterior, and the width of the posterior around it, which gives the uncertainties. Although the loss is defined over the full parameter space, its gradient and curvature describe its local behaviour at a given point. While the former indicates the direction in which the loss decreases, the latter describes how rapidly it rises away from the minimum and, hence, how tightly the data constrain each combination of parameters. An estimate of the curvature therefore serves three purposes: it sets the length and direction of each step, it tells when the minimum has been reached, and it gives the error bars (Sec.~\ref{sec:errors}). Implementation details are given in Appendix~\ref{app:optimization}.

A first-order optimizer such as Adam \citep{Kingma2014_Adam} uses only the gradient, rescaled separately for each parameter, and the length of its steps is set by a learning rate. This works well when the parameters are roughly independent, but in the orbit problem the physical parameters are strongly correlated, most notably $M$, $D$ and $a$, and a hair parameter, where present, is correlated with $M$. Along such correlated directions the appropriate step length differs by orders of magnitude, according to how tightly the data constrain each direction, and a single learning rate cannot capture that. The physics term in Eq.~\eqref{eq:loss} also makes the problem stiff. With $\epsilon=10^{-4}$, a change of the hair that alters the rate $F$ by $10^{-3}$ raises each physics residual $R_c/\epsilon$ by about 10, unless the network changes at the same time. The minimum therefore lies in a narrow, curved valley in the joint space of physical parameters and network weights, and moving along it requires a coordinated change of the hair, the other physical parameters and the network. Adam progresses only slowly along this valley. 

The curvature is available at little extra cost because the loss is a sum of squares. Let $\mathbf r$ be the vector of the normalized data residuals, frame-prior terms and the physics residuals $R_c/\epsilon$ in Eq.~\eqref{eq:loss}, so that $\mathcal L=\mathbf r^\top\mathbf r+\lVert\theta\rVert^2/\sigma_w^2$. With $J$, the Jacobian of $\mathbf r$ with respect to the physical parameters $\mathbf u$ in the whitened coordinates (of Eq.~\eqref{eq:whitening}), and to the weights $\theta$, the Hessian (matrix of second derivatives) of the loss, $\nabla^2\mathcal L$, is, 
\begin{equation}\label{eq:hessian}
  \nabla^2\mathcal L=2\Big(J^\top J+\sum_k r_k\nabla^2 r_k+D_w\Big),
\end{equation}
where $D_w$ is zero on $\mathbf u$ and the identity divided by $\sigma_w^2$ on the weights. The frame-prior terms are linear in physical parameters $\mathbf u$ so their second derivatives vanish. Near the minimum, the physics residuals are nearly zero, and the data residuals scatter with random signs, so their contributions to the second-derivative term largely cancel and it can be dropped (we quantify its effect in Appendix~\ref{app:higher}). This gives the Gauss-Newton (GN) approximation of the curvature, together with the gradient \citep{NocedalWright2006},
\begin{equation} \label{eq:H}
  H=J^\top J + D_w = \begin{pmatrix}H_{uu}&H_{u\theta}\\H_{\theta u}&H_{\theta\theta}\end{pmatrix},\quad
  \mathbf g=J^\top\mathbf r+\frac{1}{\sigma_w^2}\begin{pmatrix}\mathbf 0\\ \theta\end{pmatrix}=\begin{pmatrix}\mathbf g_u\\\mathbf g_\theta\end{pmatrix}.
\end{equation}
Notice that $H$ needs only first derivatives, which automatic differentiation provides, and it is positive semi-definite. For Gaussian data, its $\chi^2_{\rm data}$ part is the Fisher information matrix (with the network following the parameters), whose inverse is the covariance of the parameters.

With $H$ and $\mathbf g$, a small change of the parameters and weights, $\Delta\equiv(\Delta\mathbf u,\Delta\theta)$, changes the loss to approximately
\begin{equation}\label{eq:quad}
  \mathcal L(\Delta)\simeq\mathcal L(0)+2\,\mathbf g^\top\Delta+\Delta^\top H\Delta \ .
\end{equation}
Setting its derivative to zero defines the step, $\Delta=-H^{-1}\mathbf g$, at which the quadratic reaches its minimum, $\mathcal L(\Delta)=\mathcal L(0)-\mathbf g^\top H^{-1}\mathbf g$. The step is long along directions that the data constrain weakly and short along those they constrain tightly; it follows correlated directions, and it needs no learning rate. Its predicted decrease, $\mathbf g^\top H^{-1}\mathbf g$, is a change of $\chi^2$. Since a change of $\Delta\chi^2=1$ corresponds to one standard deviation, a predicted decrease $d^2$ means that the minimum is still about $d\sigma$ away, which gives a natural stopping criterion.

The quadratic model, Eq.~\eqref{eq:quad}, follows from linearizing the residuals, $\mathbf r(\boldsymbol\Delta)\simeq\mathbf r+J\boldsymbol\Delta$, and is exact only if the residuals depend linearly on the parameters, that is, if $J$ is the same everywhere. Far from the minimum, it can be poor, which is why the GN steps are taken only after an Adam stage (see below). At the minimum, terms of higher order would make the posterior skewed or bend the degeneracies. We tested this by carrying the expansion of the residuals to second order (Appendix~\ref{app:higher}). The Gaussian reproduces the means and is slightly conservative in the widths; only for a hair that enters the metric through its square is the posterior noticeably non-Gaussian (Sec.~\ref{sec:discussion}).

We therefore proceed in two stages. Starting from a Keplerian orbit determined from the astrometry, Adam first trains all parameters and weights jointly and brings them close to the minimum; it is robust to a random initialization of the network and tolerates the random shift of the collocation points. GN steps then complete the minimization in the physical parameters. Minimizing Eq.~\eqref{eq:quad} over the weights first (Appendix~\ref{app:gn}) gives the step
\begin{equation}\label{eq:gn}
\begin{split}
    \Delta\mathbf u &= - H_m^{-1}\mathbf g_m,\\
    H_m &= H_{uu} - H_{u\theta} H_{\theta\theta}^{-1} H_{\theta u},\qquad
    \mathbf g_m = \mathbf g_u - H_{u\theta}H_{\theta\theta}^{-1}\mathbf g_\theta .
\end{split}
\end{equation}
Here, $H_m$ is the curvature of the loss in the physical parameters after the network has adjusted to them. The step therefore already accounts for the change of the network needed to keep satisfying the geodesic equation, and it moves the hair, the mass and the orbital elements together along the valley. This is the coordinated move that Adam does not make.

After the step, the old weights no longer solve the geodesic equation for the new orbit. Instead of using the linear estimate of their change from the quadratic model, which is accurate only for small steps, we hold the hair at its new value and re-train the network and the other physical parameters with Adam in a short run, after which the next step is computed\footnote{Each component of $\Delta\mathbf u$ is limited to 5 whitened units (about $5\sigma$). The subsequent step is halved if the hair changes sign or if the predicted decrease grows, and a hair at the edge of its prior is held there if the step points outwards.}. We stop when the predicted decrease $\mathbf g_m^\top H_m^{-1}\mathbf g_m$ falls below $0.02$, that is, when the minimum is located to about $0.14\sigma$. Both stages minimize the same loss of the same model and differ only in how the parameters are updated. Each optimization is repeated from different starting values of the hair and, for Schwarzschild, from three initializations of the network. We accept an estimate only if the final values agree to better than $0.2\sigma$ (Appendix~\ref{app:validation}).

\subsection{Uncertainties}\label{sec:errors}

In the Laplace approximation \citep{MacKay1992}, the posterior is a Gaussian centered on the minimum of the loss, $\mathbf u^\ast$. The network weights are not physical parameters. For a given $\mathbf u$, they are set by minimizing the loss, that is, mainly by the geodesic equation, and when $\mathbf u$ changes, they follow through their linear response $A\equiv d\theta/d\mathbf u$ (Appendix~\ref{app:gn}). Along this response, the data and frame-prior residuals $\mathbf r_{\rm d}$ change by $\hat J\,\Delta\mathbf u$, with $\hat J=d\mathbf r_{\rm d}/d\mathbf u=J_u+J_\theta A$, where $J_u$ and $J_\theta$ are the rows of $J$ for $\mathbf r_{\rm d}$. The posterior of the physical parameters is therefore
\begin{equation}\label{eq:cov}
  p(\mathbf u)\propto\exp\!\Big[-\tfrac12\big(\chi^2_{\rm data}+\chi^2_{\rm frame}\big)\Big]
\simeq\exp\!\Big(-\tfrac12\,\Delta\mathbf u^\top\hat J^\top\hat J\,\Delta\mathbf u\Big),
\end{equation}
with covariance matrix $\Sigma_u=\big(\hat J^\top\hat J\big)^{-1}$ and $\Delta\mathbf u=\mathbf u-\mathbf u^\ast$.

Because the network follows the parameters, any trade-off between the network and the physical parameters is included in the error bars. Moreover, only the data and the frame priors measure $\mathbf u$: the physics residuals and the weight prior determine how the network follows, but they do not enter $\Sigma_u$. Freezing the network, or integrating over the weights as if they were parameters, would violate one of these properties and underestimate the uncertainties (Appendix~\ref{app:gn}). With Eq.~\eqref{eq:cov}, GINN reproduces the uncertainties of the exact solver, which has no network (Appendix~\ref{app:validation}).

We draw samples from $\mathcal N(\mathbf u^\ast,\Sigma_u)$ and map them to the physical parameters through Eq.~\eqref{eq:whitening}, which preserves nonlinear correlations such as the $M$-$D$ degeneracy. The same Gaussian gives the Bayesian evidence of each spacetime (Appendix~\ref{app:evidence}). Departures from it are examined in Appendix~\ref{app:higher}.

A hair parameter that enters the metric only through $\xi^2$ ($q_{\rm RN}$, $q_B$ and $q_{\rm JNW}$) needs a slightly different treatment, because at $\xi=0$ the residuals do not change to first order and the GN curvature in $\xi$ vanishes. To leading order $\chi^2_{\rm data}$ is instead linear in $\xi^2$, $\chi^2\simeq\chi^2_0+s_\xi\,\xi^2$, so the posterior is even in $\xi$ and approximately Gaussian with width $\sigma_\xi=s_\xi^{-1/2}$. We compute the slope $s_\xi=\partial\chi^2_{\rm data}/\partial(\xi^2)$ from the data residuals, with the network following the equation of motion (Appendix~\ref{app:xi2}).

For the Yukawa correction, $\delta$ and $\lambda$ are degenerate (Sec.~\ref{sec:spacetimes}). We therefore fix $\lambda$, obtain the posterior of $\delta$ with all other parameters marginalized, and repeat this on a grid of $\lambda$ to map the constraint on $\delta$ as a function of the range.

%%%%%%%%%%%%%%%%%%
\section{Results}\label{sec:results}

\subsection{Inferred parameters}

\begin{figure*}
    \centering
    \includegraphics[width=0.99\linewidth]{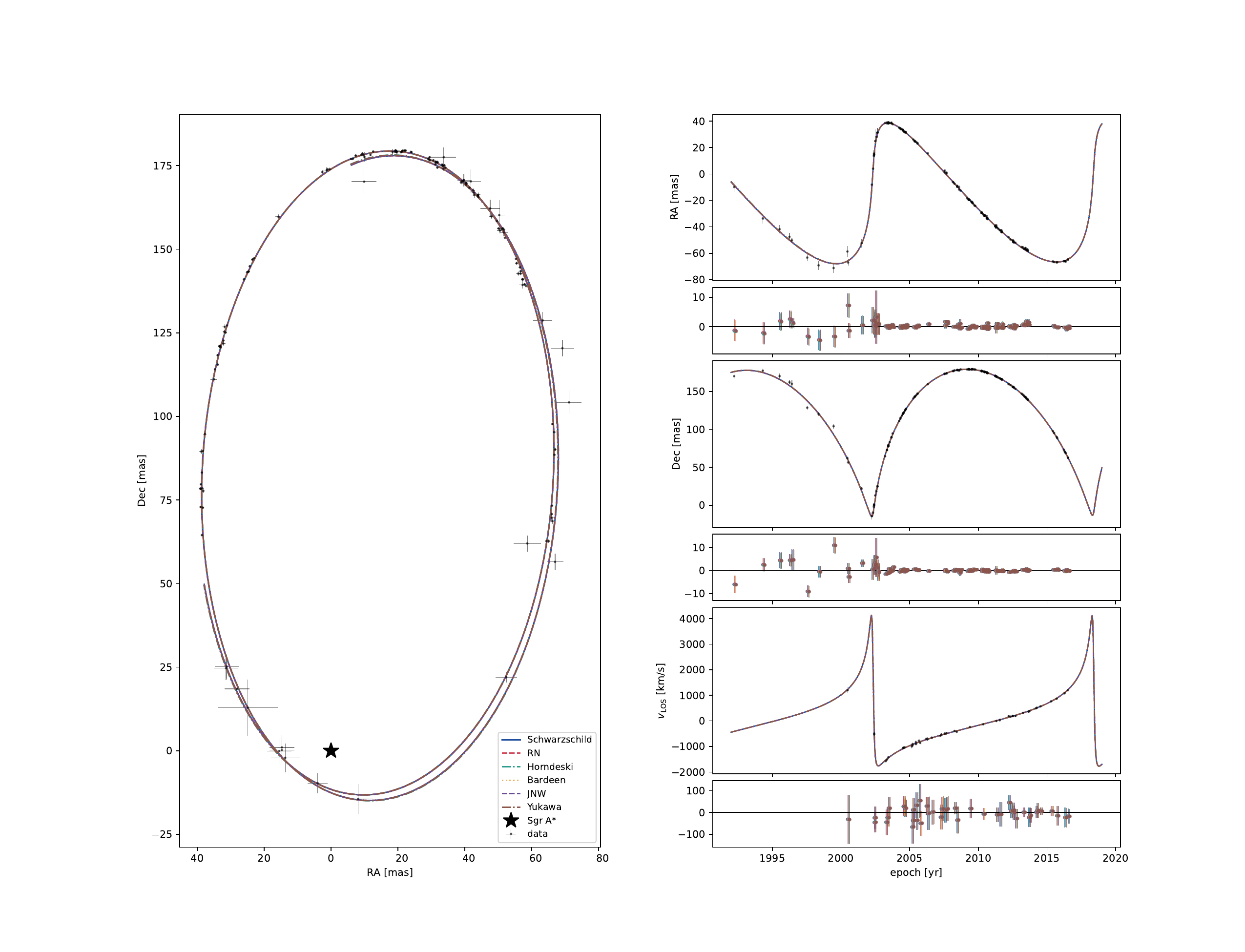}
    \caption{Best-fit orbits of S2 in all spacetimes. Left: positions on the sky relative to Sgr~A$^*$ (denoted by a star), with the astrometric data. Right: RA, Dec, and line-of-sight velocity as functions of time, each with the residuals of every spacetime below it. The orbits are almost indistinguishable at the precision of the data.}
    \label{fig:allorbits}
\end{figure*}

Table~\ref{tab:results} lists the estimates and their $1\sigma$ uncertainties, and Fig.~\ref{fig:allorbits} shows the best-fit orbits. The posteriors are shown in Fig.~\ref{fig:corner_01_schwarz} for Schwarzschild and in Appendix~\ref{app:corners} for the other spacetimes. For Schwarzschild, $\chi^2_{\rm data}=302.9$ for $N=334$ data points and $\chi^2_{\rm frame}=4.4$, so that $\chi^2_{\rm min}=307.3$ for 338 residuals and 13 parameters ($\chi^2/\nu=0.95$), with $M=(4.32\pm0.22)\times10^6\,M_\odot$ and $D=8.33\pm0.21$ kpc. These values are consistent with \citet{Gillessen2016data} and with more recent determinations \citep[e.g.,][]{GRAVITYSchprecess, GRAVITY2021DMdistribution1, GRAVITY2024extended}. In all other spacetimes, the orbital elements agree with the Schwarzschild ones to within $1\sigma$. The largest differences occur for Horndeski and Yukawa, which are the two spacetimes where the hair changes the Newtonian potential, with their masses larger by $0.2\sigma$. %In these two spacetimes the hair is also correlated with the proper motion of the frame, whose uncertainty in $v_{x,0}$ grows by 50\%. 

\begin{table*}[t]
\centering
\caption{Estimates and $1\sigma$ uncertainties from the Gauss-Newton minimum and the Laplace approximation, with the network following the parameters. The hair parameters and their priors are listed in Table~\ref{tab:priors}. For $q_{\rm RN}$, $q_B$ and $q_{\rm JNW}$, which enter the metric only through their square, the sign of the estimate has no meaning, and the width is computed as in Sec.~\ref{sec:errors}; their 95\% upper limits on $|\xi|$ are given in Sec.~\ref{sec:results}. The Yukawa results are for $\lambda=10^3$\,AU.}
\label{tab:results}
\small
\setlength{\tabcolsep}{3pt}
\begin{tabular}{lcccccc}
\hline\hline
Parameter & Schwarzschild & Hairy Horndeski & Reissner-Nordstr\"om & JNW & Bardeen & Yukawa \\
\hline
$M \; [10^6\,M_\odot]$ & $4.32 \pm 0.22$ & $4.36 \pm 0.24$ & $4.32 \pm 0.22$ & $4.32 \pm 0.22$ & $4.32 \pm 0.22$ & $4.36 \pm 0.23$ \\
$D \; [\mathrm{kpc}]$ & $8.33 \pm 0.21$ & $8.32 \pm 0.21$ & $8.32 \pm 0.21$ & $8.33 \pm 0.21$ & $8.32 \pm 0.21$ & $8.35 \pm 0.21$ \\
\hline
Hair $\xi$ & - & $q_H$ & $q_{\rm RN}$ & $q_{\rm JNW}$ & $q_B$ & $\delta$ \\
& - & $0.0019 \pm 0.0035$ & $0.33 \pm 2.53$ & $0.9 \pm 128$ & $3.0 \pm 70.4$ & $0.0047 \pm 0.0119$ \\
\hline
$a \; [\mathrm{mas}]$ & $124.58 \pm 0.96$ & $124.61 \pm 0.96$ & $124.60 \pm 0.96$ & $124.58 \pm 0.97$ & $124.58 \pm 0.96$ & $124.51 \pm 0.97$ \\
$e$ & $0.8821 \pm 0.0022$ & $0.8817 \pm 0.0024$ & $0.8822 \pm 0.0022$ & $0.8821 \pm 0.0022$ & $0.8822 \pm 0.0022$ & $0.8818 \pm 0.0025$ \\
$i \; [^\circ]$ & $134.80 \pm 0.45$ & $134.78 \pm 0.45$ & $134.79 \pm 0.45$ & $134.80 \pm 0.45$ & $134.80 \pm 0.45$ & $134.83 \pm 0.45$ \\
$\Omega \; [^\circ]$ & $227.13 \pm 0.71$ & $227.08 \pm 0.72$ & $227.14 \pm 0.71$ & $227.13 \pm 0.71$ & $227.13 \pm 0.71$ & $227.06 \pm 0.74$ \\
$\omega \; [^\circ]$ & $65.25 \pm 0.69$ & $65.11 \pm 0.75$ & $65.26 \pm 0.69$ & $65.25 \pm 0.69$ & $65.25 \pm 0.69$ & $65.10 \pm 0.79$ \\
$t_p-2002 \; [\mathrm{yr}]$ & $0.3210 \pm 0.0060$ & $0.3202 \pm 0.0066$ & $0.3211 \pm 0.0060$ & $0.3210 \pm 0.0060$ & $0.3210 \pm 0.0060$ & $0.3200 \pm 0.0068$ \\
\hline
$x_0 \; [\mathrm{mas}]$ & $0.130 \pm 0.136$ & $0.114 \pm 0.139$ & $0.130 \pm 0.137$ & $0.129 \pm 0.136$ & $0.130 \pm 0.137$ & $0.120 \pm 0.140$ \\
$y_0 \; [\mathrm{mas}]$ & $-0.278 \pm 0.191$ & $-0.284 \pm 0.191$ & $-0.279 \pm 0.191$ & $-0.279 \pm 0.192$ & $-0.279 \pm 0.192$ & $-0.286 \pm 0.192$ \\
$v_{x,0} \; [\mathrm{mas\,yr^{-1}}]$ & $0.108 \pm 0.041$ & $0.132 \pm 0.060$ & $0.109 \pm 0.041$ & $0.108 \pm 0.041$ & $0.108 \pm 0.041$ & $0.124 \pm 0.060$ \\
$v_{y,0} \; [\mathrm{mas\,yr^{-1}}]$ & $0.105 \pm 0.054$ & $0.113 \pm 0.056$ & $0.105 \pm 0.054$ & $0.105 \pm 0.054$ & $0.105 \pm 0.054$ & $0.109 \pm 0.056$ \\
$v_{z,0} \; [\mathrm{km\,s^{-1}}]$ & $5.9 \pm 7.5$ & $6.7 \pm 7.7$ & $5.8 \pm 7.5$ & $5.9 \pm 7.5$ & $5.9 \pm 7.5$ & $6.7 \pm 7.8$ \\
\hline\hline
\end{tabular}
\end{table*}

None of the hair parameters differs significantly from its GR value. The Horndeski hair, $q_H=(1.9\pm3.5)\times10^{-3}$, is consistent with zero. The Reissner-Nordstr\"om charge, $q_{\rm RN}=0.3\pm2.5$, does not separate the black-hole range $|q_{\rm RN}|\le1$ from a naked singularity, and is best read as a bound on a positive $1/r^2$ term in the metric function $f(r)$. For Bardeen and JNW, whose corrections at the radii of S2 are suppressed by $\sim10^{-7}\xi^2$, the data exclude only $|\xi| \gtrsim 10^2$ with $q_B = 3\pm70$ and $q_{\rm JNW} = 1\pm128$. These values lie far outside the horizon and photon-sphere ranges of both spacetimes, so S2 and horizon-scale imaging probe different parts of their parameter space. Our conclusions agree with the MCMC analyses of the same spacetimes by \citet{testingbhspacetimebayesian} and \citet{Bambhaniya:2026slo}, who also find no preference over Schwarzschild.

For the three hair parameters that enter only through $\xi^2$, the posterior is even in $\xi$. The best-fit values differ from zero only within the stopping tolerance of the optimization ($0.14\sigma$). We therefore summarize these posteriors by upper limits on $|\xi|$. Since the posterior of these hairs is not Gaussian, we take the limits from the higher-order posterior (Sec.~\ref{sec:discussion}): $|q_{\rm RN}|<3.9$, $|q_B|<92$ and $|q_{\rm JNW}|<175$ (95\%), in agreement with the exact solver to within 2\%. The Gaussian of Table~\ref{tab:results} gives the weaker limits 5.0, 138 and 251.

At fixed $\lambda=10^3$\,AU, comparable to the semi-major axis of S2 ($a\simeq1030$\,AU), we find $\delta=0.005\pm0.012$, consistent with zero at $0.4\sigma$. To map the constraint as a function of the range, we use 15 values of $\lambda$ between 10 and $3\times10^4$\,AU, logarithmically spaced. At each value, all other parameters and the network are re-optimized with GN steps, starting from the solution at the neighbouring $\lambda$. The posterior of $\delta$, a Gaussian $\mathcal N(\delta^\ast,\sigma_\delta^2)$ restricted to the prior, gives the 95\% upper limit $\delta_{95}$ on $|\delta|$ through $P(|\delta|<\delta_{95})=0.95$. 

\begin{figure}
  \centering
  \includegraphics[width=0.99\columnwidth]{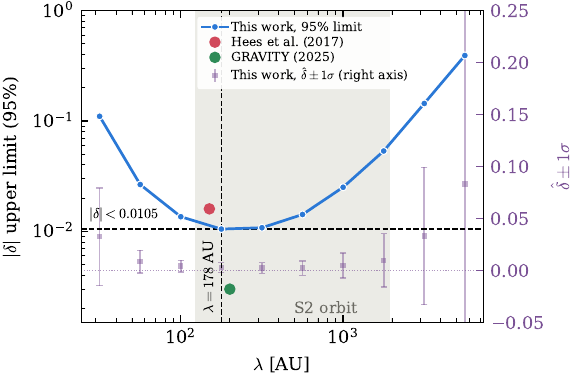}
  \caption{Yukawa correction as a function of its range $\lambda$. Blue (left axis): 95\% upper limit on $|\delta|$; the dashed lines mark the tightest limit. Purple (right axis): estimate $\delta^\ast$ and its $1\sigma$ uncertainty. The grey band marks the radial range of S2, from pericenter to apocenter. Red and green points are the tightest limits of \citet{Hees2017GRtests} and \citet{GRAVITYfifth}.}
  \label{fig:yukawa}
\end{figure}

Figure~\ref{fig:yukawa} shows the result. The limit is tightest at $\lambda\simeq180$\,AU ($2.7\times10^{13}$\,m), with $|\delta|<0.0105$ and $\delta^\ast=0.0027\pm0.0047$. For $\lambda$ much smaller than the pericenter distance, the Yukawa term has decayed along the whole orbit and only rescales the mass. For $\lambda$ much larger than the orbit, it reduces to a constant and a uniform force. The limit therefore weakens on both sides, to $|\delta|<0.11$ at $\lambda\simeq32$\,AU and $|\delta|<0.14$ at $\lambda\simeq3\times10^3$\,AU, and reaches the prior boundary for $\lambda\lesssim18$\,AU and $\lambda\gtrsim10^4$\,AU. The estimate $\delta^\ast$ is consistent with zero at all $\lambda$. Our tightest limit is comparable to that of \citet{Hees2017GRtests}, $|\alpha|<0.016$ at $\lambda=150$\,AU\footnote{The potential we use, $-2M(1+\delta e^{-r/\lambda})/[(1+\delta)r]$, has the same radial dependence as theirs, $-2M'(1+\alpha e^{-r/\lambda})/r$ with $M'=M/(1+\delta)$, so $\delta=\alpha$. The two parametrizations differ only in which mass is called $M$.}, and is about 3.5 times weaker than that of \citet{GRAVITYfifth}, $|\alpha|<0.003$ at $\lambda\simeq200$\,AU, which includes GRAVITY astrometry up to 2022. Adding four more stars and data up to 2025, \citet{GRAVITY2026Yukawa} recently tightened this to $|\alpha|<6\times10^{-4}$ at the same range.
 
\subsection{Model comparison}
 
The Laplace approximation also gives the integral of the posterior, so the evidence of each spacetime follows from quantities already computed at the minimum. Relative to Schwarzschild, the logarithm of the Bayes factor is (Appendix~\ref{app:evidence})
\begin{equation}\label{eq:lnB}
  \ln B=-\frac{\Delta\chi^2_{\rm min}}{2}+\ln\frac{\sqrt{2\pi}\,\sigma_\xi\,m_\xi}{\Delta_\xi}+\frac12\,\Delta\ln\det\Sigma_{\rm rest}.
\end{equation}
Here $\Delta\chi^2_{\rm min}$ is the change of the minimum, $\Delta_\xi$ the width of the hair prior, $m_\xi$ the part of the Gaussian $\mathcal N(\xi^\ast,\sigma_\xi^2)$ inside it, and $\Sigma_{\rm rest}$ the covariance of the other parameters at fixed hair. The second term is the Occam factor, the fraction of the prior range of the hair allowed by the data. It penalizes a hair that the data do not require, and it depends on the chosen prior. The last term is below 0.02 for all spacetimes.

Table~\ref{tab:evidence} lists the values. All the Bayes factors are negative, so none of the spacetimes is preferred over Schwarzschild. On the scale of \citet{Trotta2008} ($|\ln B|<1$ inconclusive, $1$-$2.5$ weak, $2.5$-$5$ moderate), the evidence for Schwarzschild ranges from inconclusive to moderate. In every case, the Bayes factor is set by the Occam penalty of the hair rather than by the fit. Horndeski and Yukawa lower $\chi^2_{\rm min}$ by 0.3 and 0.2, less than needed to pay for the prior range of their hair, and the other three spacetimes do not improve the fit at all. For Reissner-Nordstr\"om, $\ln B$ is therefore entirely the penalty of the prior range $|q_{\rm RN}|\le50$. Restricted to the black-hole range $|q_{\rm RN}|\le1$, narrower than the posterior, the prior would give $\ln B\simeq0$. For hair that enters the metric through its square, the posterior is narrower than its Gaussian, and using the higher-order posterior changes $\ln B$ by less than 0.1, which does not change these conclusions (Appendix~\ref{app:higher}).

\begin{table}
  \centering
  \caption{Laplace estimate of the Bayes factor relative to Schwarzschild, Eq.~\eqref{eq:lnB}, with the Occam factor $\ln O=\ln[\sqrt{2\pi}\,\sigma_\xi m_\xi/\Delta_\xi]$. $\ln B>0$ would favour the spacetime over Schwarzschild.}
  \label{tab:evidence}
  \begin{tabular}{lccc}
    \hline\hline
    Spacetime & $\Delta\chi^2_{\rm min}$ & $\ln O$ & $\ln B$ \\
    \hline
    Hairy Horndeski             & $-0.31$ & $-2.45$ & $-2.29$ \\
    Reissner-Nordstr\"om        & $0.01$  & $-2.76$ & $-2.77$ \\
    Bardeen                     & $0.00$  & $-0.82$ & $-0.82$ \\
    JNW                         & $0.00$  & $-1.32$ & $-1.32$ \\
    Yukawa ($\lambda=10^3$\,AU) & $-0.18$ & $-3.51$ & $-3.41$ \\
    \hline
  \end{tabular}
\end{table}

\begin{figure*}
    \centering
    \includegraphics[width=\linewidth]{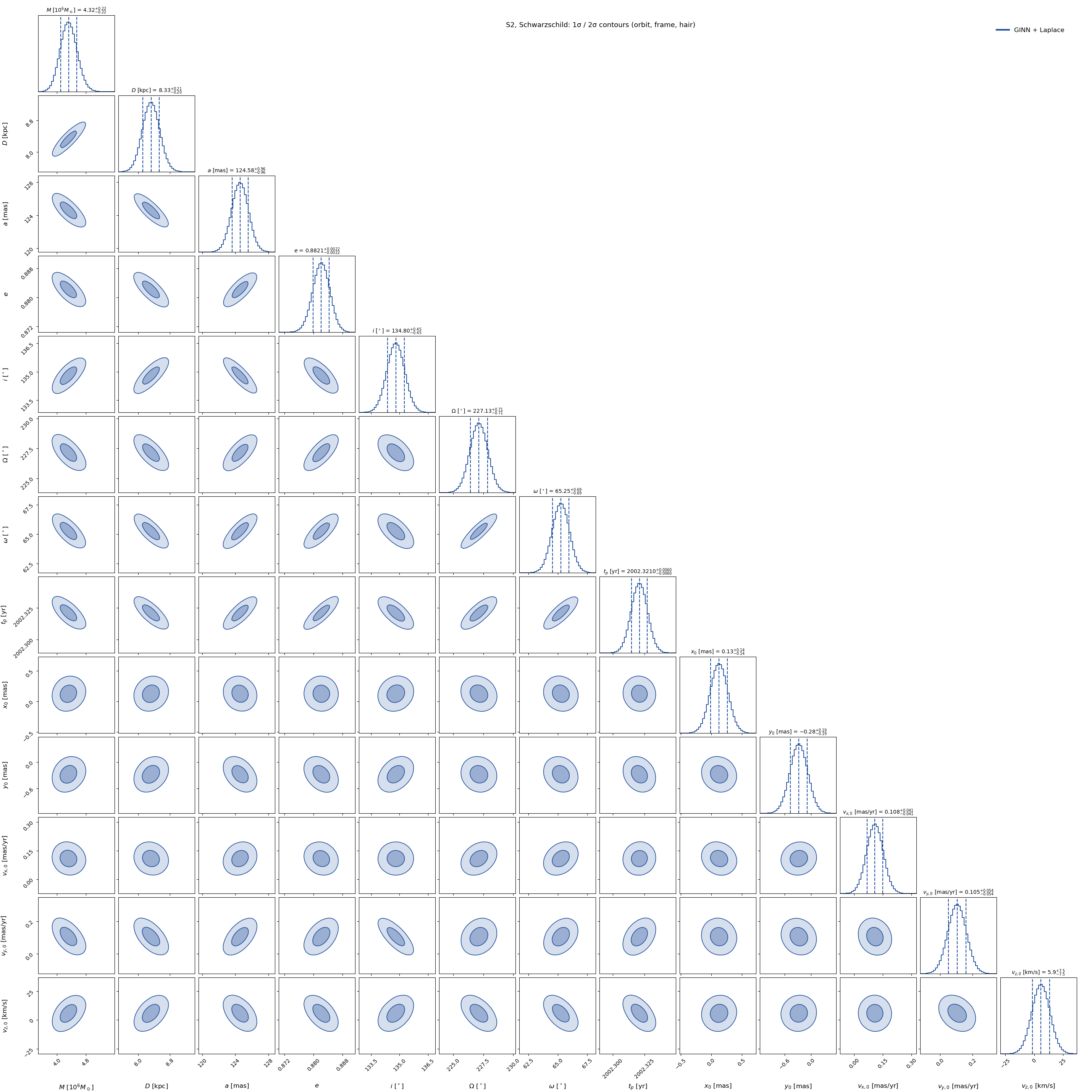}
    \caption{Posterior of all 13 parameters for Schwarzschild, from the Laplace approximation, with the network following the parameters. The contours enclose 39\% and 86\% of the two-dimensional probability ($1\sigma$ and $2\sigma$). In the titles we show the median and the 16th and 84th percentiles.}
    \label{fig:corner_01_schwarz}
\end{figure*}

\section{Discussion}\label{sec:discussion}

The results rest on three considerations: the network replaces the exact solution of the geodesic equation, the optimizer finds the minimum of the loss, and the posterior is Gaussian. We test each of them separately (Appendices~\ref{app:higher} and \ref{app:validation}). 

\begin{figure*}
    \centering
    \includegraphics[width=0.99\linewidth]{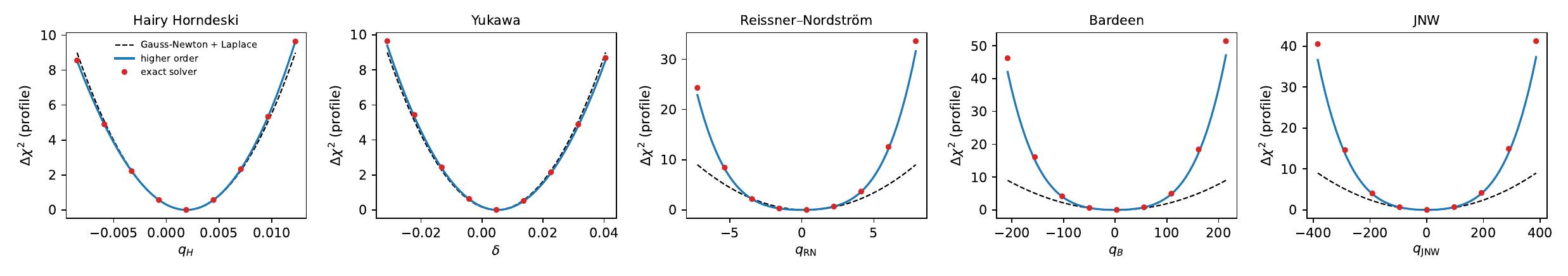}
    \caption{Profile $\Delta\chi^2$ for each hair parameter (minimized over all others), comparing the Gaussian (Laplace) approximation (dashed), the second-order expansion (solid), and the exact solver (points). For linear hairs ($q_H$, $\delta$) all three agree closely; for squared hairs ($q_{\rm RN}$, $q_B$, $q_{\rm JNW}$), $\chi^2$ rises faster than a parabola, yielding a narrower posterior than the Gaussian. For these, the parameter that the data constrain is $\xi^2$ (Appendix~\ref{app:xi2}).}
    \label{fig:profiles}
\end{figure*}

\begin{itemize}
    \item \textit{Accuracy of the network}: The decisive test is a comparison with an exact geodesic solver, which inverts $t(\phi)$ numerically and contains neither the network nor the physics term. For Schwarzschild, the estimates of the two models differ by less than $0.01\sigma$ and their uncertainties by less than $1\%$. The comparison also holds away from the minimum; over the whole posterior of every spacetime, the exact solver and GINN assign nearly the same probability to the same parameters. The network is therefore as accurate as the exact solver wherever the posterior has weight. Letting the network follow the parameters nevertheless matters for the uncertainties: with the weights held fixed, they can be up to $20\%$ smaller, because the response of the network to the parameters partly compensates the change in the Keplerian timing. 

    \item \textit{Convergence:} Repeating each optimization from different starting values of the hair, and for Schwarzschild from different initializations of the network, gives final values of the hair, $M$ and $D$ that agree to within $0.15\sigma$ in all spacetimes. The Adam stage alone does not achieve this, and the GN stage is what makes the result independent of the starting point. 

    \item \textit{Beyond the Gaussian approximation}: A Gaussian built from the local curvature is fast, but it can misrepresent both the size and the orientation of the posterior when the observables depend nonlinearly on the parameters. We test it in two steps (Appendix~\ref{app:higher}). First, we expand the residuals to second order in the parameters, as in the DALI method of \citet{Sellentin2014}, with the network eliminated to the same order. The resulting $\chi^2$ remains a sum of squares and can describe skewed and curved posteriors. Second, we reweigh samples of this expansion with the exact solver, which gives the exact posterior.

    The six spacetimes fall into two groups. For Schwarzschild, Horndeski and Yukawa, the Gaussian describes the posterior well. It reproduces the means of all parameters to within about $0.1\sigma$ of the exact posterior and the widths of $q_H$ and $\delta$ to within about 1\%, and it overestimates the widths of the orbital elements by by 6-12\%. The reason is the second-derivative term $\sum_k r_k\nabla^2r_k$ dropped in Eq.~\eqref{eq:hessian}; for these data it does not cancel along the direction in which all orbital elements are correlated, and steepens $\chi^2$ along it by 25\% (Appendix~\ref{app:higher}). For Reissner-Nordstr\"om, Bardeen and JNW, whose hair enters only through $\xi^2$, the posterior of the hair is not Gaussian. Its $\chi^2$ also contains a term in $\xi^4$, and the other parameters compensate it along a curved degeneracy, so that $\chi^2$ rises faster than the parabola at large $|\xi|$ (Fig.~\ref{fig:profiles}). The exact posterior can then be up to 30\% narrower than the Gaussian, and for these three spacetimes we take the limits on $|\xi|$ from the higher-order posterior (Sec.~\ref{sec:results}). In neither group does the Gaussian underestimate the uncertainties by more than a per cent, and in both the higher-order posterior follows from derivatives of the same model.
    %The posteriors of $M$ and $a$ are slightly skewed, with a skewness below 0.3.
\end{itemize}

\section{Conclusions}\label{sec:conclusion}

We have presented the Geodesic-Informed Neural Network (GINN), a forward model for stellar orbits that is exact in the metric and differentiable in all parameters. A small network, constrained by the geodesic equation, represents only the radial phase $\phi(t)$. A single optimization that includes Gauss-Newton steps gives the estimates, and the curvature at the minimum, with the network following the parameters, gives their uncertainties and the Bayesian evidence.

Applied to the 1992-2016 data of S2, GINN reproduces an exact geodesic solver and finds no significant deviation from Schwarzschild in any of the six spacetimes considered. The tightest constraints are on hair that modifies the Newtonian potential, and the Bayes factors favour Schwarzschild mainly through the Occam penalty of the additional parameter. For Schwarzschild, Horndeski and Yukawa, whose hair enters the metric linearly, the Gaussian describes the posterior well. For Reissner-Nordstr\"om, Bardeen and JNW, whose hair enters only through its square, the Gaussian is conservative, and the limits on the hair follow from a second-order expansion of the same model, checked against the exact solver.

The method has limitations. The Laplace approximation and its higher-order extension describe the posterior around one maximum and would not reveal a separate mode; the multi-start test makes such a mode unlikely but does not exclude it. The forward model assumes a static, spherically symmetric metric and test-particle motion, so a rotating black hole would require a different parametrization of the orbit. Our data do not include GRAVITY astrometry, and we do not model astrometric confusion \citep{PlewaSari2018}.

Because a spacetime enters only through $f$, $g$ and $h$, GINN applies to any static, spherically symmetric metric without modification. It can be extended directly to several stars sharing one metric, each with its own network, and to GRAVITY data, for which the non-Keplerian correction becomes measurable. Finally, the same construction can represent the metric functions themselves by networks, turning the test of given spacetimes into a reconstruction of the spacetime probed by the orbits, which will be discussed in a future work.

\begin{acknowledgements}
      PHP is funded by  Fundação de Amparo à Pesquisa e Inovação do Espírito Santo (FAPES, Brazil) and Coordenação de Aperfeiçoamento de Pessoal de Nível Superior (CAPES, Brazil). We acknowledge the use of computational resources from the \href{https://computacaocientifica.ufes.br/scicom}{Sci-Com Lab} of the Department of Physics at UFES, supported by FAPES, CAPES, and CNPq. DF, AB and DFM thank the Research Council of Norway for their support and the resources provided by UNINETT Sigma2 -- the National Infrastructure for High Performance Computing and Data Storage in Norway.
\end{acknowledgements}

% 
% Please note that we have included the references to the file aa.dem in
% order to compile it, but we ask you to:
%
% - use BibTeX with the regular commands:
\bibliographystyle{aa} % style aa.bst
\bibliography{biblography} % your references Yourfile.bib
%
% - join the .bib files when you upload your source files
%-------------------------------------------------------------------

%#####################
\newpage
\clearpage
\pagebreak

\appendix
\section{Optimization details}\label{app:optimization}

\subsection{Optimization steps}

\begin{enumerate}
    \item \textit{Initialization:} We initialize the parameters with a Keplerian fit to the astrometry alone. For a fixed period, eccentricity and time of pericenter $(P,e,t_p)$, the projected orbit is linear in the Thiele-Innes constants, which follow from weighted least squares. $(P,e,t_p)$ are found by a grid search refined with the Nelder-Mead method \citep{Nelder1965}, and the constants are converted to $(a,i,\Omega,\omega)$. The mass follows from Kepler's third law at a nominal distance. Astrometry alone cannot separate $M$ from $D$, so this value is only a starting guess, and $D$ is determined together with $M$ afterwards. The network output is scaled by $10^{-3}$\,rad, so the initial correction to the Keplerian orbit is small. The hair starts slightly away from its GR value; for a hair entering only through $\xi^2$, the gradient would otherwise vanish.
    
    \item \textit{Whitened coordinates:} The physical parameters differ in scale by orders of magnitude and are correlated. Adam works in whitened coordinates $\mathbf u$,
    \begin{equation}\label{eq:whitening}
        \Theta = \Theta_0 + S\, W \mathbf u,\qquad W W^\top = H_0^{-1},
    \end{equation}
    where $\Theta_0$ is the current point and $S$ a diagonal matrix of rough parameter scales. $H_0$ is the GN matrix of $\chi^2_{\rm data} + \chi^2_{\rm frame}$ in the scaled variables $S^{-1}(\Theta-\Theta_0)$, computed with the network switched off and with a small ridge to keep it invertible. $W$ is the lower-triangular Cholesky factor of $H_0^{-1}$. Since $W^\top H_0W=I$, the components of $\mathbf u$ are uncorrelated with unit variance, and a step of given size corresponds to about the same number of standard deviations in every direction. The whitening is recomputed at the start of every Adam run, including the re-training between GN steps. It is only a change of variables: the loss, its minimum, the GN step and the covariance mapped back to $\Theta$ do not depend on it.
    
    \item \textit{Adam optimization:} In the first stage Adam runs for $N_A=3000$ steps with learning rates $\eta_0=0.1$ for $\mathbf u$ and $5\times10^{-4}$ for the weights. A larger rate for the weights makes Adam oscillate, because the physics term is stiff. Both rates follow a cosine schedule \citep{Loshchilov2017},
    \begin{equation}
        \eta_k=\eta_{\min}+\frac{(\eta_0-\eta_{\min})}{2}\left[1+\cos\left(\frac{\pi k}{K}\right)\right],\quad \eta_{\min}=0.01\,\eta_0 .
    \end{equation}
    Because the collocation points are shifted randomly at each step, the physics term fluctuates, and with a constant rate, Adam keeps moving around the minimum by an amount proportional to the rate. The decreasing rate lets it settle. We keep the state with the lowest loss on a fixed collocation grid, so a rate that decays before convergence cannot leave a worse state behind. Final convergence in the physical parameters is in any case reached in the GN stage. Each re-training between GN steps is a short Adam run of 400 steps with the same schedule, with the hair held fixed and a learning rate of $2\times10^{-3}$ for the weights, since the network then only has to adapt to a small change of the orbit. The GN iteration also stops after 30 steps or after more than five halvings of the step, but in all our fits the criterion on the predicted decrease was met first.
\end{enumerate}

\subsection{Reduced step and Laplace posterior}\label{app:gn}
In block form, Eq.~\eqref{eq:quad} reads
\begin{equation}\label{eq:quadblock}
\begin{split}
    \mathcal L(\Delta) &\simeq\mathcal L(0) + 2\big(\mathbf g_u^\top\Delta\mathbf u+\mathbf g_\theta^\top\Delta\theta\big) + \Delta\mathbf u^\top H_{uu}\Delta\mathbf u\\
    &\quad + 2\,\Delta\mathbf u^\top H_{u\theta}\Delta\theta+\Delta\theta^\top H_{\theta\theta}\Delta\theta ,
\end{split}
\end{equation}
where the weight prior $D_w$ makes $H_{\theta\theta}$ invertible. Completing the square in $\Delta\theta$ gives
\begin{equation}\label{eq:square}
\begin{split}
  \mathcal L(\Delta)\simeq\mathcal L(0)-\mathbf g_\theta^\top H_{\theta\theta}^{-1}\mathbf g_\theta
  &+\big(\Delta\theta-\Delta\theta^\ast\big)^\top H_{\theta\theta}\big(\Delta\theta-\Delta\theta^\ast\big)\\
  &+2\,\mathbf g_m^\top\Delta\mathbf u+\Delta\mathbf u^\top H_m\Delta\mathbf u ,
\end{split}
\end{equation}
with $\Delta\theta^\ast=-H_{\theta\theta}^{-1}(\mathbf g_\theta+H_{\theta u}\Delta\mathbf u)$, and with $H_m$ and $\mathbf g_m$ as in Eq.~\eqref{eq:gn}. This one identity gives both the step and the response of the network that enters the covariance.

\begin{enumerate}
    \item \textit{Step:} For any $\Delta\mathbf u$, the loss is lowest when the network takes its linear response $\Delta\theta=\Delta\theta^\ast$. What remains is a quadratic in $\Delta\mathbf u$ with curvature $H_m$ and gradient $\mathbf g_m$, which is minimized by the step of Eq.~\eqref{eq:gn}. The total predicted decrease is $\mathbf g_\theta^\top H_{\theta\theta}^{-1}\mathbf g_\theta+\mathbf g_m^\top H_m^{-1}\mathbf g_m$. The first term is the gain from adjusting the network alone. It is nearly zero after re-training, since then $\mathbf g_\theta\simeq0$, so only the second term is used to stop.

    \item \textit{Laplace posterior:} At the minimum $\mathbf g_u=\mathbf g_\theta=0$, so the response of the network to a displacement $\Delta\mathbf u$ is $\Delta\theta^\ast=A\,\Delta\mathbf u$ with $A=-H_{\theta\theta}^{-1}H_{\theta u}$. We take this response as the solution of the equation of motion for the new parameters. Along it, the data and frame-prior residuals change by $\hat J\Delta\mathbf u$, with $\hat J=J_u+J_\theta A$ restricted to these rows, and the posterior of the physical parameters is $p(\Delta\mathbf u)\propto\exp(-\frac12|\hat J\Delta\mathbf u|^2)$, a Gaussian with covariance $\Sigma_u=(\hat J^\top\hat J)^{-1}$, Eq.~\eqref{eq:cov}.     The two other ways of treating the network both underestimate the uncertainties:
    \begin{itemize}
      \item \textit{Network fixed}, $\Delta\theta=0$, gives $\Sigma_u=(J_u^\top J_u)^{-1}$. This ignores that a change of the orbit changes the solution of the geodesic equation, so the trade-off between the network and the physical parameters is missing. For example, the uncertainty of $M$ for Schwarzschild is then about 20\% too small ($0.175$ instead of $0.22$).
      \item \textit{Network integrated over}: integrating the Gaussian of Eq.~\eqref{eq:square} over $\Delta\theta$ gives $\Sigma_u=H_m^{-1}$, the $\mathbf u$ block of $H^{-1}$. This treats the physics residuals and the weight prior as measurements of $\mathbf u$, although they only fix how the network represents the solution. For example, the uncertainty of the Yukawa parameter $\delta$ is then about 15\% too small.
    \end{itemize}
    The GN step uses $H_m$, because it minimizes the full loss, including the physics term; the covariance uses $\hat J^\top\hat J$, because only the data and the frame priors are measurements.
\end{enumerate}
 
\subsection{Hair parameters entering as $\xi^2$}\label{app:xi2}
 
Let the residuals depend on the hair only through $\xi^2$. By the chain rule, $\partial r_k/\partial\xi=2\xi\,\partial r_k/\partial(\xi^2)$, which vanishes at $\xi=0$, and so does the GN curvature $2\sum_k(\partial r_k/\partial\xi)^2$. The curvature of $\chi^2_{\rm data}=\sum_kr_k^2$ comes entirely from the term dropped in Eq.~\eqref{eq:hessian}. With $\partial^2r_k/\partial\xi^2=2\,\partial r_k/\partial(\xi^2)$ at $\xi=0$,
\begin{equation}
  \frac{\partial^2\chi^2}{\partial\xi^2}\bigg|_{\xi=0}=2\sum_k r_k\frac{\partial^2r_k}{\partial\xi^2}=2\,\frac{\partial\chi^2}{\partial(\xi^2)}\equiv2\,s_\xi .
\end{equation}
To leading order in the hair, the residuals are linear in $\xi^2$, so $\chi^2\simeq\chi^2_0+s_\xi\,\xi^2$. Since the posterior is $\propto e^{-\chi^2/2}$, it is approximately a Gaussian of width $\sigma_\xi=s_\xi^{-1/2}$, provided $s_\xi>0$. It is even in $\xi$, so at $\xi=0$ the hair is uncorrelated with the other parameters, which keep the covariance $\Sigma_u$ of Eq.~\eqref{eq:cov}.
 
The slope cannot be read off a Jacobian at $\xi=0$, where the Jacobian vanishes. We therefore evaluate it at a finite value $\xi_e$, where $\chi^2$ is still linear in $\xi^2$. The hair is set to $\xi_e$ and the network is re-trained. The data residuals are then linearized in all free parameters, with the network following through the linear response that keeps the physics residuals fixed. Minimizing the resulting quadratic model over the other parameters gives the slope $\partial\chi^2/\partial\xi$ at $\xi_e$, and $s_\xi$ is this slope divided by $2\xi_e$. %Only the data residuals enter, because the physics residuals constrain the network and are not measurements of $\xi$.
 
This is the same rule as for a hair that enters linearly. In both cases $\sigma_\xi$ is the shift of the hair from its best-fit value at which the $\chi^2$, minimized over all other parameters, has risen by one. For a linear hair, this $\chi^2$ is quadratic in $\xi$, which gives the Laplace width; for a squared hair it is linear in $\xi^2$, which gives $s_\xi^{-1/2}$. The next-order terms make the squared-hair posterior somewhat narrower than this Gaussian (Appendix~\ref{app:higher}). The parameter that the data constrain is therefore $\eta=\xi^2$, in which the residuals are linear to first order. Fitting $\eta$ instead of $\xi$ would make the Gaussian more accurate, but a prior uniform in $\xi$ is not uniform in $\eta$, and the limits in the literature are quoted for $|\xi|$, so we keep $\xi$.

\section{Bayesian evidence}\label{app:evidence}
 
The evidence of a model with parameters $\Theta$ and prior $\pi(\Theta)$ is $Z=\int\mathcal P(\mathbf y\,|\,\Theta)\,\pi(\Theta)\,d\Theta$. The Laplace approximation replaces the integrand by a Gaussian around the maximum, which gives
\begin{equation}
  Z\simeq\mathcal P(\mathbf y\,|\,\Theta^\ast)\,\pi(\Theta^\ast)\,(2\pi)^{n/2}\det\Sigma^{1/2},
\end{equation}
where $\mathcal P\propto e^{-\chi^2/2}$, $\Sigma$ is the Laplace covariance of Sec.~\ref{sec:errors} expressed in $\Theta$, and $n$ is the number of parameters. All spacetimes share the same 13 orbital and frame parameters. The determinant factorizes as $\det\Sigma=\sigma_\xi^2\det\Sigma_{\rm rest}$, with $\sigma_\xi^2$ the marginal variance of the hair and $\Sigma_{\rm rest}$ the covariance of the other parameters at fixed hair. A flat hair prior of width $\Delta_\xi$ contributes $1/\Delta_\xi$, and restricting the Gaussian to the prior contributes the fraction $m_\xi$ inside it. Together these give Eq.~\eqref{eq:lnB}.
 
We make three assumptions here: (i) only $\chi^2_{\rm data}$ and $\chi^2_{\rm frame}$ enter $\Delta\chi^2_{\rm min}$ and $\Sigma$; the normalization of the frame priors is the same in all spacetimes and cancels. The physics residuals are left out because at the minimum they vanish to within the tolerance, and since the network follows the parameters rather than being integrated over (Sec.~\ref{sec:errors}), it contributes no factor of its own to the evidence. (ii) the Occam term depends on the prior through $-\ln\Delta_\xi$, so the Bayes factors are meaningful only together with Table~\ref{tab:priors}. (iii) for squared hair the widths of Appendix~\ref{app:xi2} are used, and the remaining non-Gaussianity changes $\ln B$ by less than 0.1, without changing the conclusions. We compute this change by importance sampling the higher-order posterior of Appendix~\ref{app:higher}, with a proposal built from its samples.

\section{Beyond the quadratic approximation}\label{app:higher}

In coordinates $\mathbf z$ in which the Laplace posterior is $\mathcal N(0,I)$, $\mathbf u=\mathbf u^\ast+C\mathbf z$ with $CC^\top=\Sigma_u$, we expand the data residuals to second order,
\begin{equation}\label{eq:dali}
  \mathbf r(\mathbf z)\simeq\mathbf r_0+J\mathbf z+\tfrac12K(\mathbf z,\mathbf z),
\end{equation}
with $J_{k,i}=\partial r_k/\partial z_i$ and $K_{k,ij}=\partial^2 r_k/\partial z_i\partial z_j$. Squaring gives
\begin{multline}\label{eq:dali_chi2}
  \chi^2(\mathbf z)=|\mathbf r_0|^2+2\,\mathbf r_0^\top J\mathbf z+\mathbf z^\top\big(J^\top J+\mathbf r_0\!\cdot\!K\big)\mathbf z\\
  +(J\mathbf z)^\top K(\mathbf z,\mathbf z)+\tfrac14\,|K(\mathbf z,\mathbf z)|^2,
\end{multline}
with $(\mathbf r_0\!\cdot\!K)_{ij}=\sum_k r_{0,k}K_{k,ij}$. Keeping only $J$ gives back the Laplace Gaussian. The term $\mathbf r_0\!\cdot\!K$ is the curvature dropped in Eq.~\eqref{eq:hessian}, the cubic term skews the posterior, and the quartic term bends its degeneracies. The gradient $\mathbf r_0^\top J$ vanishes at the minimum, up to the stopping tolerance, so the expansion changes the shape of the posterior but not its maximum. For $\mathbf r_0=0$ and without the network, Eq.~\eqref{eq:dali_chi2} is the DALI doublet of \citet{Sellentin2014} (also see \citep{deSouza:2025qok}); the terms in $\mathbf r_0$ arise because we expand at the best fit to real data rather than at a fiducial model. As in DALI, the truncation is in the order of the derivatives of the residuals, so that $\chi^2$ remains a sum of squares and $e^{-\chi^2/2}$ a proper distribution.
 
When the physical parameters change, the network must keep solving the geodesic equation. We eliminate its weights to the same order, $\theta(\mathbf z)=\theta^\ast+A\mathbf z+\frac12B(\mathbf z,\mathbf z)$. Let $\tilde J_\theta$ be the Jacobian of all residuals (data, physics and weight prior) with respect to the weights, and $P=\tilde J_\theta(\tilde J_\theta^\top\tilde J_\theta)^{-1}\tilde J_\theta^\top$ the projector onto the changes that the network can produce. The first-order response is that of Appendix~\ref{app:gn}, $A=-H_{\theta\theta}^{-1}H_{\theta z}$, and turns the Jacobian into $(I-P)J_z$, whose data and frame-prior rows give $J^\top J=I$ in these coordinates. At second order, minimizing $|K(\mathbf z,\mathbf z)|^2$ over $B$ gives $B_{ij}=-(\tilde J_\theta^\top\tilde J_\theta)^{-1}\tilde J_\theta^\top K^{\rm lin}_{ij}$, where $K^{\rm lin}$ is the second derivative along $\theta^\ast+A\mathbf z$, so that $K=(I-P)\,K^{\rm lin}$. Only the data and frame-prior rows enter Eq.~\eqref{eq:dali}, since the physics residuals are the constraint that the network follows rather than measurements, as for the covariance, Eq.~\eqref{eq:cov}. $J$ and $K^{\rm lin}$ are computed by automatic differentiation.

Since the expansion leaves the maximum in place, its effect is on integrals over the posterior, which we obtain by sampling $e^{-\chi^2(\mathbf z)/2}$, restricted to the hair prior, with \texttt{emcee} \citep{ForemanMackey2013}. The polynomial $\chi^2$ is cheap to evaluate, and we use $4n$ walkers for $n$ parameters and run each chain until it is longer than 50 autocorrelation times and gives at least $10^4$ independent samples, discarding the first five autocorrelation times. The chains are longer than 180 autocorrelation times, so that their Monte Carlo errors are well below the effects we report. To obtain the exact posterior, we reweight 6000 of these samples by $w\propto\exp[-\frac12(\Delta\chi^2_{\rm exact}-\Delta\chi^2)]$, where $\Delta$ denotes the change from the minimum and $\chi^2_{\rm exact}$ is computed with the exact solver. The effective fraction of samples, $(\sum_i w_i)^2/(n_s\sum_i w_i^2)$ for $n_s$ samples, measures the agreement: it equals 1 when the two posteriors coincide.
 
The exact solver keeps 86-100\% of the samples effective in all spacetimes, so the expansion, and with it the network, describes the posterior well. The term $\mathbf r_0\!\cdot\!K$ explains the narrowing of the orbital elements. For Schwarzschild, the eigenvalues of $J^\top J+\mathbf r_0\!\cdot\!K$ lie within 2\% of one, except one of 1.25. Its eigenvector is correlated by 0.8-0.99 with each orbital element, so it is the direction that sets their uncertainties, and narrowing it by $1.25^{-1/2}\simeq0.90$ reduces their widths by 6-10\%, as found. The orbital elements narrow by the same amount in all spacetimes. Over noise realizations $\mathbf r_0\!\cdot\!K$ averages to zero, but for one data set it need not be small. For a hair entering as $\xi^2$, Eq.~\eqref{eq:dali} gives along $\xi$, with $\mathbf c=\partial\mathbf r/\partial(\xi^2)$, $\chi^2=|\mathbf r_0+\mathbf c\,\xi^2|^2=\chi^2_0+s_\xi\,\xi^2+|\mathbf c|^2\xi^4$. Here $s_\xi=2\,\mathbf r_0^\top\mathbf c$ gives the width of Appendix~\ref{app:xi2}, and the quartic term, together with a curved degeneracy with the other parameters, makes the posterior narrower.

\section{Validation}\label{app:validation}

\begin{itemize}
    \item \textit{Exact solver}: The exact solver obtains $\phi(t)$ by Newton inversion of $t(\phi)=\int F^{-1}d\phi$ and is otherwise identical to GINN, with the same observables and optimizer but without the network and the physics term. For Schwarzschild, the estimates of both models differ by less than $0.01\sigma$, and their uncertainties by less than 1\%. The network reproduces the exact non-Keplerian correction to 0.4\% (rms), and the improvement of the fit over a Keplerian orbit is the same for both, $\Delta\chi^2_{\rm data}=0.127$ with the network and $0.126$ with the exact solver.

    \item \textit{Physics tolerance}: Repeating the Schwarzschild optimization for $\epsilon$ between $3\times10^{-2}$ and $10^{-5}$ gives three regimes. For $\epsilon\gtrsim3\times10^{-3}$ the network is only loosely bound to the equation of motion and absorbs noise, so that the minimum $\chi^2$ falls below the exact-solver value and $M$ and $D$ are biased. For $10^{-4}\lesssim\epsilon\lesssim10^{-3}$ the results agree with the exact solver and do not depend on $\epsilon$. For $\epsilon\lesssim3\times10^{-5}$ the problem becomes too stiff to converge in the same number of steps. We use $\epsilon=10^{-4}$.

    \item \textit{Multi-start}: Each optimization is repeated from different starting values of the hair spanning the prior or several orders of magnitude, and for Schwarzschild from different initializations of the network. The final values are stable: across all spacetimes, the largest spread in the final values of the hair, $M$, and $D$ over three starts is below $0.04\sigma$ (except for Hairy Horndeski spacetime where they agree within $0.15\sigma$). 
    
    \item \textit{Exact posterior}: Furthermore, replacing the Gaussian approximation with the exact non-Gaussian posterior shifts the means by at most $0.1\sigma$ in all cases. The non-Gaussianity mainly affects the widths of the squared-hair parameters: the exact posteriors for $q_{\rm RN}$, $q_B$, and $q_{\rm JNW}$ have standard deviations $0.82$, $0.70$, and $0.72$ times those of the Gaussian, the orbital elements 0.89-0.94 times, and $q_H$ and $\delta$ 1.01 times. The exact posterior reweights between 86\% (Bardeen, JNW) and 100\% (Schwarzschild, Horndeski, Yukawa) of the higher-order samples effectively.
\end{itemize}

\section{Posteriors of the other spacetimes}\label{app:corners}
 
\begin{figure}
    \centering
    \includegraphics[width=\linewidth]{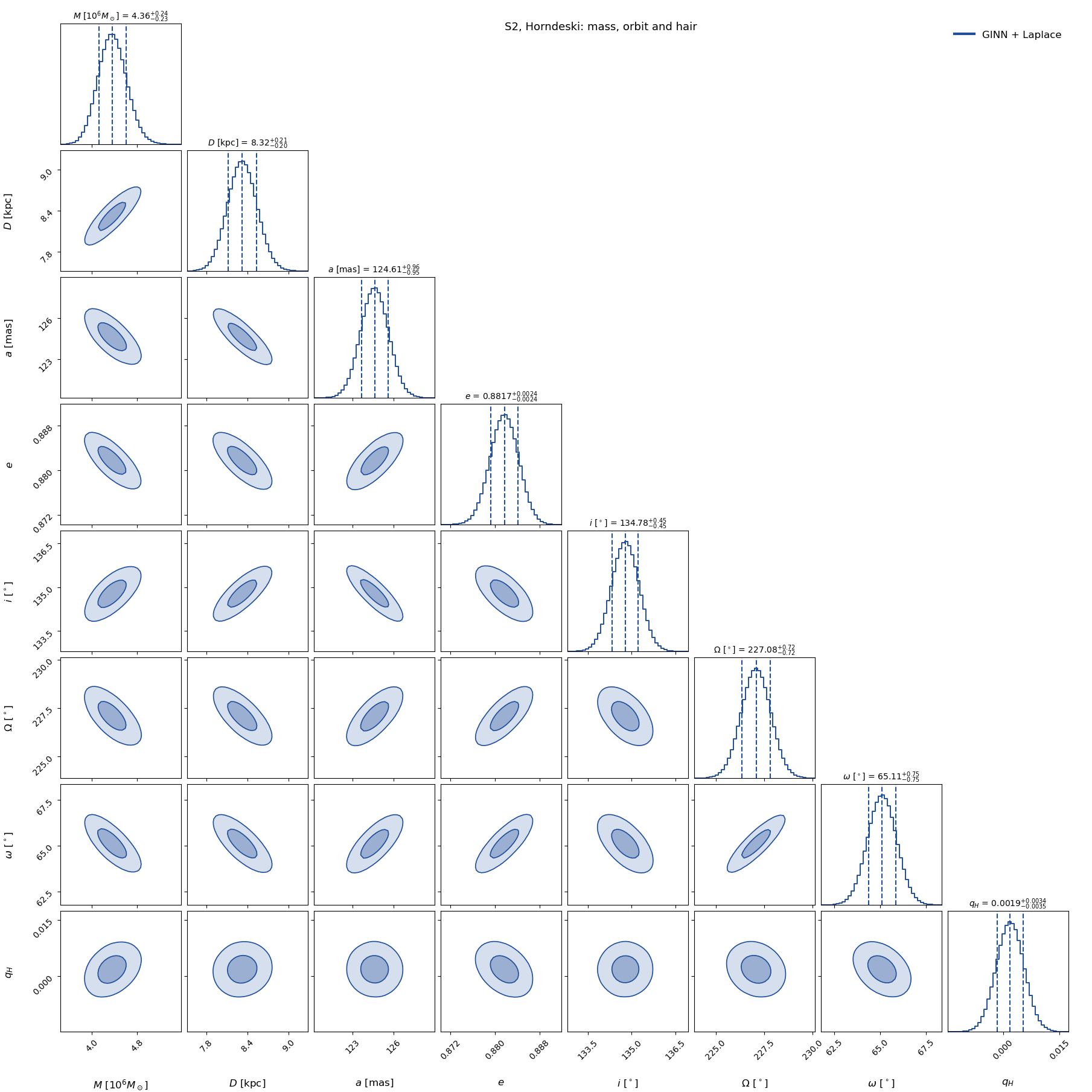}
    \caption{Posterior for the hairy Horndeski spacetime.}
    \label{fig:small_corner_02_hairy_horndeski}
\end{figure}

\begin{figure}
    \centering
    \includegraphics[width=\linewidth]{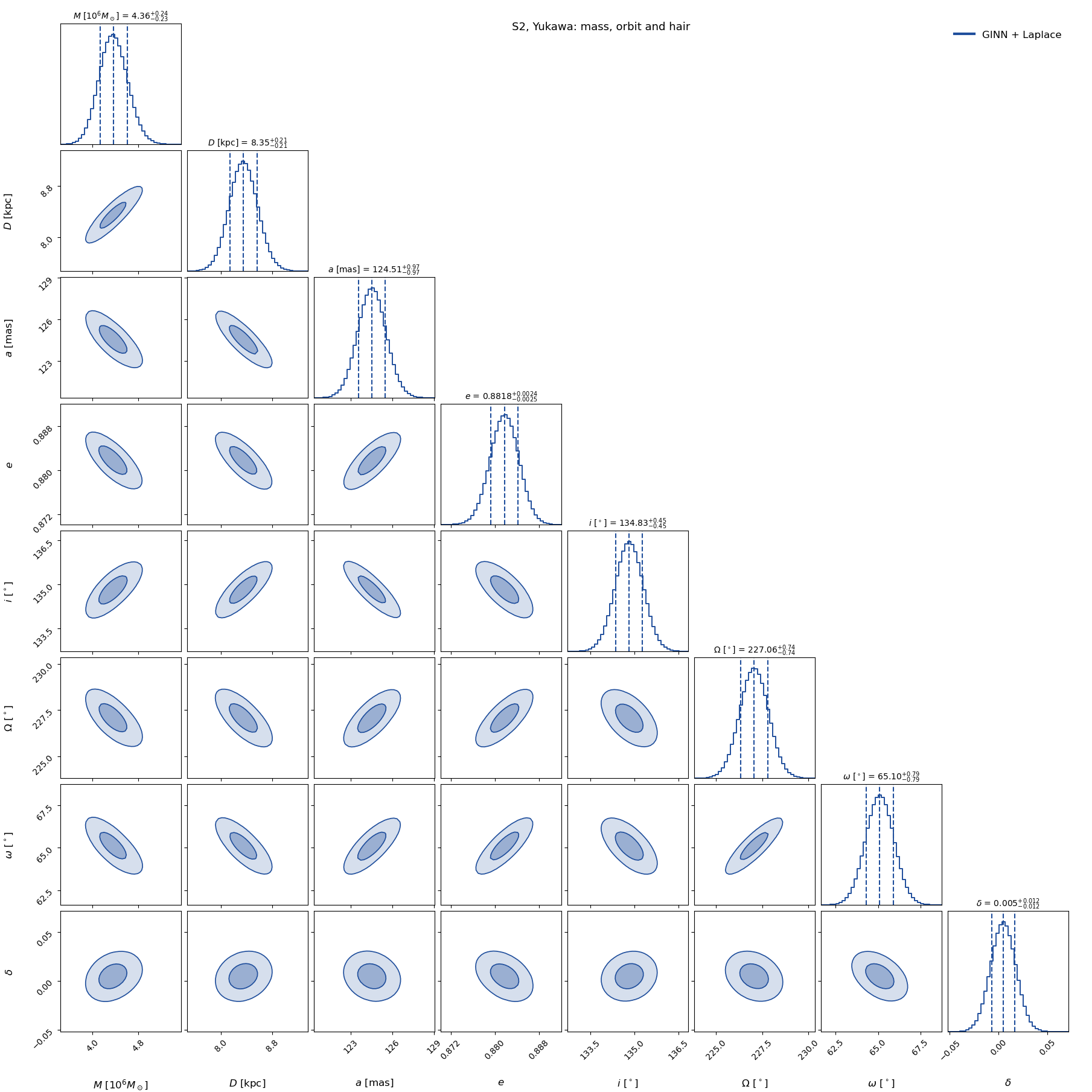}
    \caption{Posterior for the Yukawa correction at fixed $\lambda=10^3$\,AU.}
    \label{fig:small_corner_03_yukawa}
\end{figure}

\begin{figure}
    \centering
    \includegraphics[width=\linewidth]{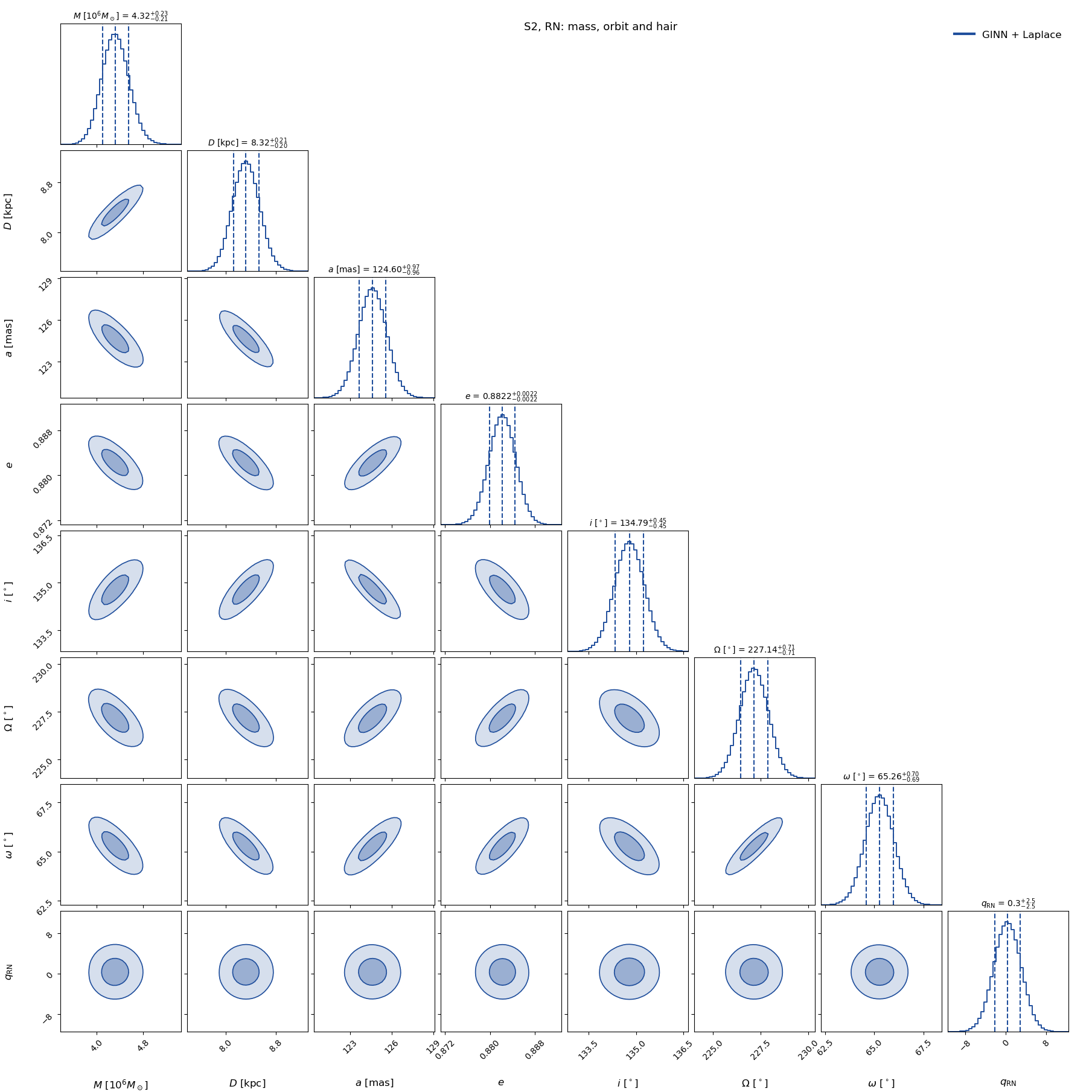}
    \caption{Posterior for the Reissner-Nordstr\"om spacetime.}
    \label{fig:small_corner_05_RN}
\end{figure}

\begin{figure}
    \centering
    \includegraphics[width=\linewidth]{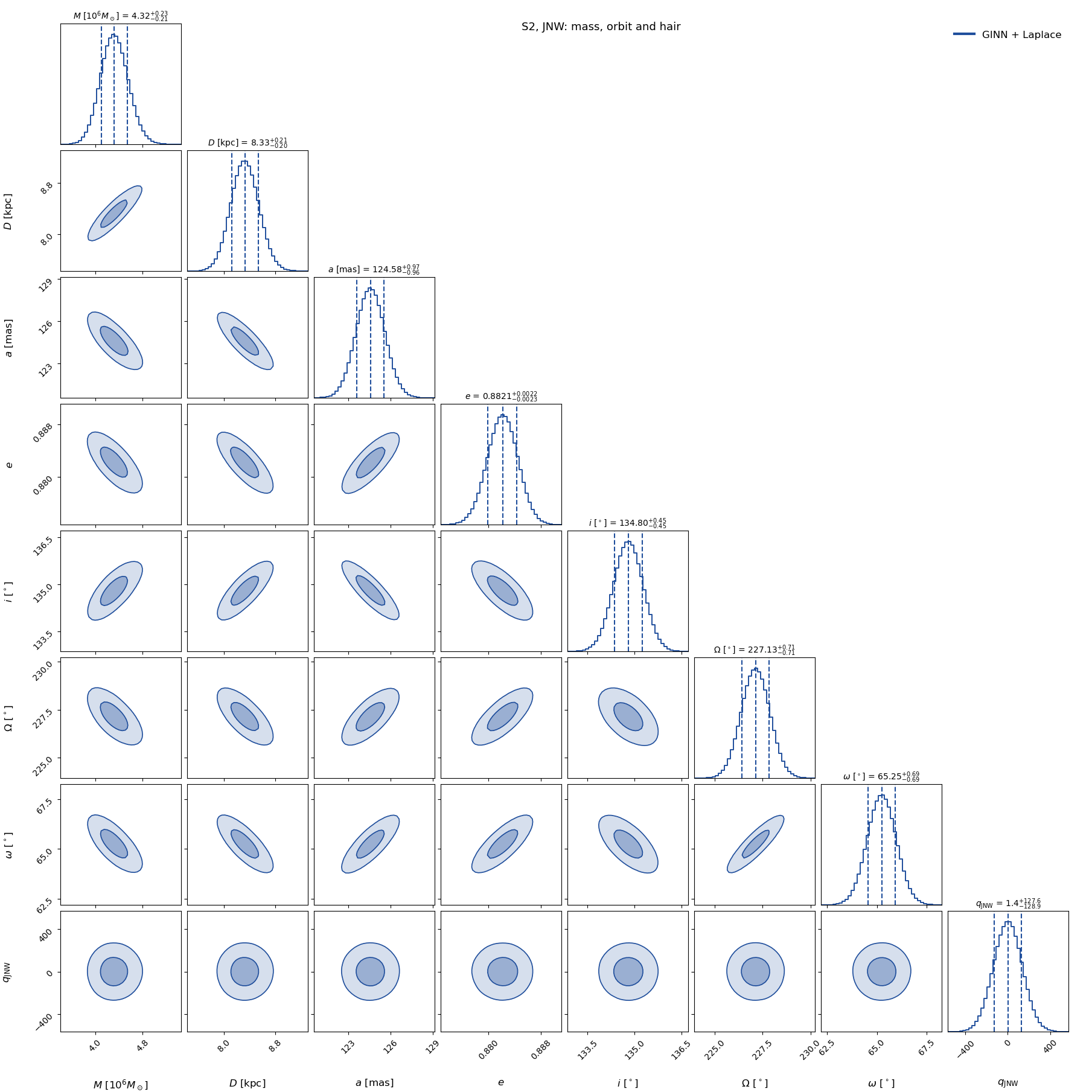}
    \caption{Posterior for the JNW spacetime.}
    \label{fig:small_corner_06_JNW}
\end{figure}

\begin{figure}
    \centering
    \includegraphics[width=\linewidth]{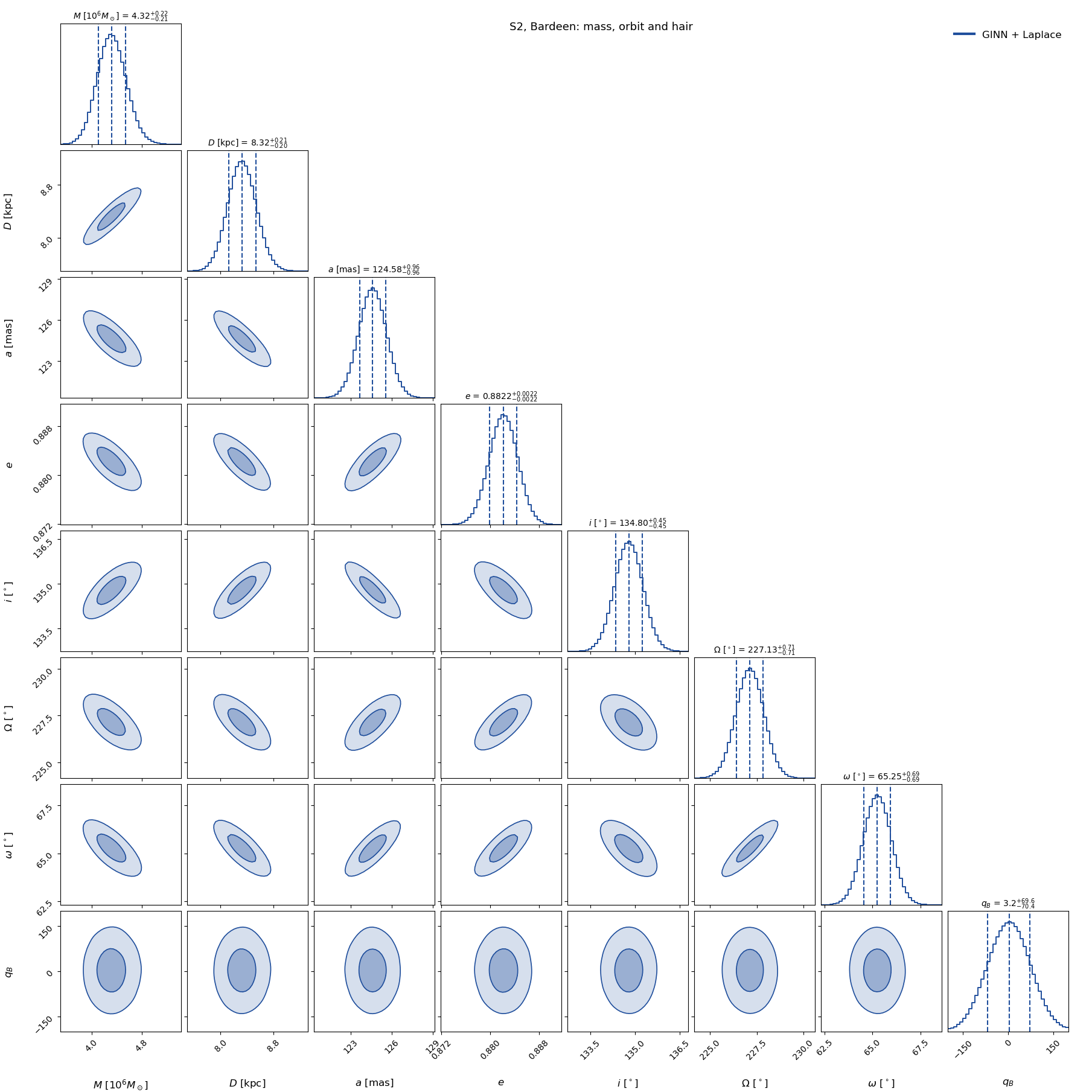}
    \caption{Posterior for the Bardeen spacetime.}
    \label{fig:small_corner_07_bardeen}
\end{figure}

Figures~\ref{fig:small_corner_02_hairy_horndeski}-\ref{fig:small_corner_07_bardeen} show the posteriors of the mass, the orbital elements and the hair for the other spacetimes, from the Laplace approximation, with the network following the parameters. The reference-frame parameters are omitted for brevity; their values are listed in Table~\ref{tab:results}. The contours enclose 39\% and 86\% of the two-dimensional probability ($1\sigma$ and $2\sigma$), and the titles give the median and the 16th and 84th percentiles.
 
\end{document}